\documentclass[fleqn,usenatbib]{mnras}

\usepackage{newtxtext,newtxmath}

\usepackage[T1]{fontenc}
\usepackage{ae,aecompl}

\usepackage{graphicx}	
\usepackage{amsmath}	
\usepackage{lscape}

\title[GPM for Faraday Depth Spectra]{Gaussian Process Modelling for Improved Resolution in Faraday Depth Reconstruction}

\author[S.~W.~Ndiritu et al.]{
S.~W.~Ndiritu,$^{1,2}$\thanks{E-mail: sndiritu@mpa-garching.mpg.de
 (SWN)}
A.~M.~M.~Scaife$^{2,3}$,
D.~L.~Tabb$^{4}$,
M.~C{\'a}rcamo$^{2,5}$ \&
J.~Hanson$^{2}$ \\
$^{1}$ Max Planck Institute for Astrophysics, Karl-Schwarzschildstra{\ss}e 1, 85748 Garching, Germany \\
$^{2}$ Jodrell Bank Centre for Astrophysics, Department of Physics and Astronomy, University of Manchester, Alan Turing Building,\\ Oxford Road, Manchester, M13 9PL, UK \\
$^{3}$ The Alan Turing Institute, Euston Road, London NW1 2DB, UK \\
$^{4}$ Division of Molecular Biology and Human Genetics, Stellenbosch University Faculty of Medicine and Health Sciences, \\ Cape Town, South Africa. \\
$^{5}$ Departamento de Ingenier\'ia Inform\'atica, Universidad de Santiago de Chile, Av. Ecuador 3659, Santiago, Chile \\
}

\date{Accepted XXX. Received YYY; in original form ZZZ}

\pubyear{2019}

\begin{document}
\label{firstpage}
\pagerange{\pageref{firstpage}--\pageref{lastpage}}
\maketitle

\begin{abstract}
The incomplete sampling of data in complex polarization measurements from radio telescopes negatively affects both the rotation measure (RM) transfer function and the Faraday depth spectra derived from these data. Such gaps in polarization data are mostly caused by flagging of radio frequency interference and their effects worsen as the percentage of missing data increases. In this paper we present a novel method for inferring missing polarization data based on Gaussian processes (GPs). Gaussian processes are stochastic processes that enable us to encode prior knowledge in our models. They also provide a comprehensive way of incorporating and quantifying uncertainties in regression modelling. In addition to providing non-parametric model estimates for missing values, we also demonstrate that Gaussian process modelling can be used for recovering rotation measure values directly from complex polarization data, and that inferring missing polarization data using this probabilistic method improves the resolution of reconstructed Faraday depth spectra.
\end{abstract}

\begin{keywords}
techniques: polarimetric -- methods: statistical -- radio continuum: general
\end{keywords}



\section{Introduction}

Polarisation measurements from radio telescopes with broadband receiver systems have changed the way in which we investigate magnetised astrophysical plasmas by enabling the use of the RM synthesis method \citep{1966MNRAS.133...67B, 2005A&A...441.1217B}. In this method the Fourier relationship between polarized intensity as a function of wavelength-squared and the Faraday dispersion function is exploited to recover the polarized intensity as a function of Faraday depth, $\phi$, such that
\begin{equation}
\label{eq:fourier}
F(\phi) = \int_{-\infty}^{\infty}{ P(\lambda^2){\rm e}^{-2i\phi \lambda^2}~{\rm d}\lambda^2 },
\end{equation}
where
\begin{equation}
\label{eq:pol}
P(\lambda^2) = |P(\lambda^2)|{\rm e}^{2i\chi(\lambda^2)} = Q(\lambda^2) + iU(\lambda^2).
\end{equation}

Although conceptually simple, in practical terms the implementation of this method and the interpretation of the resulting Faraday dispersion function are complicated by a number of factors. The first of these complications is that the Stokes pseudo-vectors $Q$ and $U$ are not sampled natively in wavelength-squared but linearly in frequency. A further complication is that Eq.~\ref{eq:fourier} does not represent a true Fourier relationship as $P(\lambda^2)$ does not exist at $\lambda^2 < 0$. Since $F(\phi)$ is not necessarily a purely real quantity this represents a fundamental limitation in attempting to reconstruct an unknown Faraday dispersion function from measured values of $P(\lambda^2)$. An additional limitation comes from the finite bandwidth and hence range of wavelength-squared, $\Delta (\lambda^2)$, measured by a radio telescope receiver, as well as the potentially incomplete sampling over this bandwidth due to various effects (RFI, instrumental problems etc.) that require flagging (excision) of specific frequency channel data during an observation.

Incomplete sampling of complex polarization in frequency-space, as well as the non-linear mapping of frequency to $\lambda^2$, results in a non-uniform distribution of measurements in $\lambda^2$-space. This is typically represented as a multiplicative weighting function, $W(\lambda^2)$, which results in the convolution of the Faraday dispersion function with a transfer function,
\begin{equation}
\label{eq:rmtf}
{\rm RMTF}(\phi) = \frac{\int_{-\infty}^{\infty} { W(\lambda^2){\rm e}^{-2i\phi\lambda^2}~{\rm d}\lambda^2 }}{\int_{-\infty}^{\infty} { W(\lambda^2)~{\rm d}\lambda^2 }},
\end{equation}
known as the rotation measure transfer function, RMTF$(\phi)$ \citep{2005A&A...441.1217B}. Consequently, the complications caused by both the $\lambda^2 < 0$ problem and the mapping of linear frequency to wavelength-squared may be encapsulated in the weighting function $W(\lambda^2)$ and its Fourier counterpart, RMTF($\phi$). Although attempts have been made to deconvolve the RMTF from the Faraday depth spectra, e.g. \cite{2009IAUS..259..591H}, the methods are inherently under-constrained due to the $P(\lambda^2<0) = \varnothing$ problem.

This  can cause statistical ambiguity in the results of such a deconvolution process, as Faraday depth is inherently complex-valued and therefore not subject to the same conjugate symmetry as, for example, interferometric imaging. It has also been shown that certain Faraday structures will themselves not have conjugate symmetry \citep{Brandenburg_2014}. Consequently, under-constrained deconvolution processes are liable to introduce spurious non-physical structures, whilst also being unable to reconstruct all true physical structures \citep{JP2012,pratley2020}.

These issues are of particular importance when complex or compound Faraday structures are seen along a single line of sight. In a blind data challenge comparing different methods for extracting the parameters of Faraday structures, \cite{Sun_2015} noted that RM synthesis was highly prone to missing composite Faraday structures and that the $\sim$25\% of extragalactic sources found observationally by \cite{law2011} to be Faraday complex was likely  an underestimate. \cite{Sun_2015} concluded that when Faraday complexity is present, only QU-fitting methods will produce results that do not introduce extraneous scatter in recovered RM values. Such QU-fitting methods, e.g. \cite{Farnsworth_2011, 10.1111/j.1365-2966.2012.20554.x, 2014ApJ...792...51I}, model the observed polarization, $\tilde{P}(\lambda^2)$, directly rather than decomposing Faraday depth spectra to recover parameters. However, whilst QU-fitting methods can be more reliable for recovering composite structures, in particular those including Faraday thick components, in general they rely on physically parameterised models of the synchrotron-emitting and Faraday rotating media along the line of sight to be known a priori. One recent exception to this is the deconvolution method proposed by \cite{pratley2020}, which uses a constrained {\sc clean}-style non-parametric QU-fitting approach that combines QU-fitting with deconvolution.

In this work we explore the potential for using a Gaussian Process Modelling (GPM) approach to improve both the interpretation of Faraday depth spectra and the recovery of physical parameters from polarization measurements. In its simplest form, a GPM approach can be considered part of the family of QU-fitting methods; however, it has certain advantages over physically parameterised QU-fitting methods, which include both a robustness to using non-exact models for imperfectly known datasets and an ability to provide predictive models for imputation including posterior variances.  The contents of this paper are as follows, in \S~\ref{sec:method} we describe the underlying GPM methodology including a description of the covariance kernel function employed and the optimization of hyper-parameters; in \S~\ref{sec:sims} we illustrate the impact of GP optimisation and re-sampling of polarization data on their resulting Faraday depth spectra in the presence of significant RFI flagging, and demonstrate how the rotation measures for Faraday thin data can be recovered from the hyper-parameters of the GP; in \S~\ref{sec:joedata} we demonstrate the method using a simulated source population with observational specifications matching that of the MeerKAT radio telescope; in \S~\ref{sec:disc} we discuss the implications of these results, compare the method to previous work in the literature and consider the use of more complex covariance kernels. In Section~\ref{sec:conclusions} we draw our conclusions.

\section{Gaussian Process Modelling for Faraday Depth Reconstruction}
\label{sec:astrogpm}

Gaussian Processes (GPs) are probabilistic non-parametric models that can be used for a variety of regression and classification problems. A Gaussian Process Model (GPM) can be used to describe a given dataset by assuming a priori that the data themselves represent a realisation of a stochastic process described by a multi-variate Gaussian distribution. Such models are non-parametric as the Gaussian process (GP) is a prior on the data itself rather than on a parametric model, i.e. the complexity of the model grows with the inclusion of additional data points. GPMs are considered to be an attractive solution when modelling systems where the degree of complexity required by a parametric model is not necessarily supported by the amount of information in the observations. Overly complex parametric models are prone to over-fitting and can lead to misinterpretation of the data. In contrast, Gaussian processes are flexible probabilistic regression models that do not require exact knowledge of the underlying physical model and can also provide posterior uncertainties that naturally reflect high uncertainty in regions of data space where there are few observational constraints.

In astrophysics, Gaussian processes have been used in various applications including in the study of photometric redshifts \citep{wa06300n}, variability in stellar light curves \citep{2018MNRAS.474.2094A}, Galactic dust emission \citep{ensslinfromert}, eclipses in cataclysmic variable stars \citep{10.1093/mnras/stw2417}, transmission spectroscopy of extrasolar planets \citep{2012MNRAS.419.2683G}, and radio velocity studies for planetary candidates \citep{10.1093/mnras/stv1428}. They have also been used extensively in cosmic microwave background analysis \citep{1987MNRAS.226..655B, PhysRevD.67.023001}, as well as for modelling instrumental systematics and calibration, e.g. \cite{2012MNRAS.419.2683G, 2014MNRAS.443.2517H, 2015ApJ...800...46B, Czekala_2015, 2015MNRAS.451..680E, 10.1093/mnras/stv1428, 2016MNRAS.459.2408A, 2016MNRAS.456L...6R, 2017MNRAS.466.4250L, mertens} and interferometric image reconstruction \citep{resolve}.

\subsection{GPM for imputation of missing polarization data}
\label{sec:imputation}

The transfer function described in Equation~\ref{eq:rmtf} is a function of the observational sampling in $\lambda^2$-space. In practice this sampling is affected not only by the characteristics of the telescope's receiver system but also by the necessary removal of frequency channels affected by radio frequency interference (RFI). Radio telescopes operating at or close to L-band (1.4\,GHz) typically experience data flagging at a level of $20-50\%$ in bandwidth due to a wide variety of both terrestrial and satellite RFI sources \citep[e.g.][]{deep2}. As well as negatively impacting the sensitivity of an observation, this flagging can also result in large contiguous gaps in the frequency coverage that have a detrimental effect on the form of the RMTF.

Contingent on there being sufficient information in the remaining data, Gaussian process modelling may provide a potential method for imputing\footnote{In statistics, \emph{imputation} is the process of replacing missing data points with substituted values such as model estimates.} these missing data subject to only weak assumptions about the form of the data and without the need for a parameterised physical model. Furthermore, GPM not only provides a model estimate of the value for each missing data point but also an uncertainty on that value, allowing astronomers to set thresholds on the degree of reliability they are prepared to accept when imputing the values of missing data points.

From the perspective of imputation in a more general sense, this kind of application addressing RFI excision can be considered as an example of recovering data missing completely at random (MCAR) and/or missing at random \citep[MAR; e.g.][]{MAR}, depending on the nature of the RFI, i.e. it is being used to repair deficits from local interruptions in the data set rather than recovering information from a signal at or below the reliable cusp of detection.

\subsection{GPM for resampling of irregular data}
\label{sec:resampling}

In addition to the imputation of missing data values, GPM can also provide a method of re-sampling observational data onto a uniform spacing in $\lambda^2$-space. Since data are natively taken in channels uniformly spaced in frequency, some previous practical implementations of RM Synthesis have employed an initial processing step to re-bin spectral data, $Q(\nu)~\&~U(\nu)$, into regularly spaced arrays of $\lambda^2$ in order to use an FFT transformation into Faraday depth space, as implemented in for example {\sc pyrmsynth}\footnote{\url{https://github.com/mrbell/pyrmsynth}} package \citep[described in][]{pratley2020a} that was used to perform RM synthesis for the LOFAR Two-metre Sky Survey \citep{lofarpyrmsynth}. Like RFI flagging, the non-linear relationship between frequency and wavelength-squared results in a distribution of measured data points that are often separated by large contiguous gaps in $\lambda^2$-space, resulting in high sidelobes in the RMTF. Furthermore, GPM can not only be used to re-sample data within the observed bandwidth but also to infer data beyond the boundaries of that bandwidth. Without additional data, these inferences will be subject to rapidly increasing degrees of uncertainty but may still be used over moderate distances in frequency space to improve the resolution of a measurement in Faraday depth. In practice, re-sampling to uniformly spaced $\lambda^2$-positions and imputing the values of data that are missing due to RFI flagging can be done simultaneously.

\section{Method}
\label{sec:method}

The basic premise of GPM is that, instead of employing a physically motivated or semi-empirical parametric model to describe a dataset, we can instead assume that the data, $\mathbf{y}(\mathbf{x})$, represent the realization of a random Gaussian process,
\begin{equation}
y(\mathbf{x}) = N(\mathbf{\mu}(\mathbf{x}), \mathbf{K}(\mathbf{x},\mathbf{x})),
\end{equation}
where $ K(\mathbf{x},\mathbf{x})$ is the matrix that defines the degree of covariance between every pair of measurement positions, $(x,x')$, for $\mathbf{x} = \{x_1, x_2, \dots, x_n\}$, and $\mathbf{\mu}(\mathbf{x})$ is an underlying mean function, which in this case is considered to be zero-valued. In the case of non-negligible polarization leakage, this would manifest as $\mu(x)\neq 0$.

The expected value of the data at all other positions, $x_{\ast}$, can be calculated as a posterior mean, $\mu_{\ast}$, with an associated posterior variance, $C_{\ast}$, such that
\begin{eqnarray}
\label{eq:postmu} \mu_{\ast} &=&  \mathbf{K}(x_{\ast},\mathbf{x})^T \mathbf{K}(\mathbf{x},\mathbf{x})^{-1} y  \\
\label{eq:postcov} C_{\ast} &=&  \mathbf{K}(x_{\ast},x_{\ast}) - \mathbf{K}(x_{\ast},\mathbf{x})^T \mathbf{K}(x_{\ast},\mathbf{x})^{-1} \mathbf{K}(\mathbf{x},x_{\ast})
\end{eqnarray}
see e.g. \cite{3569, article}.

For polarization data measured at regular intervals in frequency (spectral channels) and hence irregular intervals in $\lambda^2$, the posterior mean can be used to infer the values of the polarization data at regular intervals, $\lambda_{\ast}^2$, avoiding the need for explicit re-binning of the data, and the posterior variance can be used to define the weighting function $W(\lambda_{\ast}^2)$ at those positions and hence calculate the RMTF.

\subsection{Kernel Definition}

The covariance matrix used in Eqs.~\ref{eq:postmu}~\&~\ref{eq:postcov} can be populated analytically using a kernel function, $k$, that defines the degree of covariance for each measurement separation,

\begin{equation}
\mathbf{K}(\mathbf{x},\mathbf{x}) = \left(
\begin{array}{cccc}
k(x_1,x_1) & k(x_1,x_2) & ... & k(x_1,x_n) \\
k(x_2,x_1) & k(x_2,x_2) & ... & k(x_2,x_n) \\
\vdots & \vdots & \vdots & \vdots \\
k(x_n,x_1) & k(x_n,x_2) & ... & k(x_n,x_n)
\end{array}
\right),
\end{equation}
where $k(x_1,x_2)$ is the covariance between two measurements at a separation $|x_1 - x_2|$. We note that this assumes statistical homogeneity within a given data space, which may not be the case for data sets that combine significantly different $\lambda^2$ regimes.

%
%

\subsection{Covariance kernel definition}
\label{sec:kernels}

For the application to radio polarization data we use a compound covariance kernel. The contributions to this kernel comprise covariance arising both from the expected signal and from the properties of the measurement data. Faraday rotation results in periodicity of the complex polarization signal but the presence of additional Faraday components, including thick structures, can cause deviations from exact periodicity in the data. To represent this behaviour we use a quasi-periodic kernel formed from the multiplication of a exponential kernel and a periodic kernel.

We implement our kernel using the {\tt celerite} Gaussian processing library \citep{celerite}. Traditional GPM scales as $\mathcal{O}(N^3)$. This is due to the need to invert the covariance matrix and often makes GPM computationally intractable for large datasets. In contrast,  the {\tt celerite} GP implementation uses a restricted set of kernels that allow for fast separable evaluation with a linear complexity of $\mathcal{O}(N)$.

The {\tt celerite} kernel has a generalised exponential form,
\begin{equation}
    k(x_i,x_j) = \sum_{n=1}^{J}{\alpha_n {\rm e}^{-\beta_n |x_i - x_j|} + \alpha_n^{\ast} {\rm e}^{-\beta_n^{\ast} |x_i - x_j|}},
\end{equation}
which can be used to produce a quasi-periodic kernel of the form,
\begin{equation}
\label{eq:celkernel}
k_1(x_i,x_j)=h\,{\rm e}^{-c|x_i - x_j|}\left[\cos\left(\frac{2\pi|x_i - x_j|}{P}\right)\right].
\end{equation}
In this work we combine this quasi-periodic kernel with a white noise kernel,
\begin{equation}
\label{eq:whitekernel}
k_2(x_i,x_j)=\sigma_{\rm n}^2\,\delta_{ij},
\end{equation}
to take account of the thermal noise present on observational measurements of the complex polarization. Together, $k_1$ and $k_2$ are used to create a composite kernel
\begin{equation}
k = k_1 + k_2,
\end{equation}
where the contributions of the different components are governed by the values of their hyper-parameters, $h$, $c$, $P$ and $\sigma^2$. While the values of these hyper-parameters need to be inferred from the data, they do not explicitly characterise a functional form for the data, but instead govern the scales on which different covariance distributions act.

\subsection{Hyper-parameter Estimation}
\label{sec:parms}

The values of the hyper-parameters can be optimized by evaluation of a likelihood function using the measured data. Polarization data are complex valued, where
\begin{equation}
\mathbf{P}(\mathbf{\nu}) = \mathbf{Q}(\mathbf{\nu}) + i\mathbf{U}(\mathbf{\nu}),
\end{equation}
and so the loglikelihood function can be broken down such that
\begin{equation}
\label{eq:loglike}
\log L = \log L_{\rm Q} + \log L_{\rm U}.
\end{equation}
The components of this likelihood are
\begin{eqnarray}
\log L_{\rm Q} &=& -\frac{1}{2}(\mathbf{Q} - \mathbf{\mu}_{\rm Q})^{\rm T}\mathbf{K}_{\rm Q}^{-1}(\mathbf{Q} - \mathbf{\mu}_{\rm Q}) - \frac{1}{2}\ln |\mathbf{K}_{\rm Q}|\\
\log L_{\rm U} &=& -\frac{1}{2}(\mathbf{U} - \mathbf{\mu}_{\rm U})^{\rm T}\mathbf{K}_{\rm U}^{-1}(\mathbf{U} - \mathbf{\mu}_{\rm U}) - \frac{1}{2}\ln |\mathbf{K}_{\rm U}|
\end{eqnarray}
where,
\begin{eqnarray}
\mathbf{\mu}_{Q} &=& \mathbf{K}_{\rm Q}(x_{\ast},\mathbf{x})^T \mathbf{K}_{\rm Q}(\mathbf{x},\mathbf{x})^{-1} \mathbf{Q}\\
\mathbf{\mu}_{U} &=& \mathbf{K}_{\rm U}(x_{\ast},\mathbf{x})^T \mathbf{K}_{\rm U}(\mathbf{x},\mathbf{x})^{-1} \mathbf{U}.
\end{eqnarray}
In the case where the covariance matrix is characterised solely by stationary kernels, i.e. kernels that are functions of the sample separation, $|x_i - x_j|$, only, it can be assumed that $\mathbf{K}_{\rm U}\equiv \mathbf{K}_{\rm Q}$ and in this paper we assume a single kernel of the form given in Equation~\ref{eq:celkernel} to describe both Stokes parameters, the hyper-parameters of which are optimised using the joint likelihood in Equation~\ref{eq:loglike}.

\section{Application to Simulated Data}
\label{sec:sims}

\subsection{Example scenarios}
\label{sec:scenarios}

We illustrate the use of the GPM method on two examples: (1) a simple Faraday Thin scenario, and (2) a more complex scenario involving both Faraday Thin and Faraday Thick components. This second scenario corresponds to the example geometry outlined in \S~1 of \cite{2005A&A...441.1217B}.

\vskip .1in
\noindent
\textbf{Scenario One} In this scenario we assume polarized emission from the lobe of a radio galaxy with an intensity of 1\,Jy at a reference frequency of $\nu_{\rm ref} = 1.4$\,GHz and a synchrotron spectral index of $\alpha = 0.7$, which has a single valued Faraday depth of 50\,rad\,m$^{-2}$ such that,
\begin{eqnarray}
P_{\rm rg}(\lambda^2) &=&  \left(\frac{\lambda^2}{\lambda_{\rm ref}^2}\right)^{\alpha/2}\exp (2i\phi_1 \lambda^2) \\
&=&  \left(\frac{\lambda^2}{\lambda^2_{\rm ref}}\right)^{\alpha/2} \left[\cos (2\phi_1 \lambda^2) +  i \sin (2\phi_1 \lambda^2)\right].
\end{eqnarray}

\vskip .1in
\noindent
\textbf{Scenario Two} In this scenario we adopt the second line of sight geometry of \cite{2005A&A...441.1217B}, where
\begin{equation}
F_{\rm gal}(\phi) = \begin{cases}
(2\phi_{\rm fg})^{-1} & -\phi_{\rm fg}~<~\phi~<~\phi_{\rm fg} \\
0 & {\rm elsewhere}
\end{cases}
\end{equation}
\begin{equation}
P_{\rm gal}(\lambda^2) = \frac{\sin (2\phi_{\rm fg} \lambda^2)}{2\phi_{\rm fg}\lambda^2}
\end{equation}
and the total polarization is given by
\begin{equation}
P_{\rm tot}(\lambda^2) = P_{\rm gal}(\lambda^2) + P_{\rm rg}(\lambda^2).
\end{equation}

In both scenarios we assume that the signal is measured over a frequency bandwidth from 0.58\,GHz to 2.50\,GHz in 512 evenly spaced frequency channels. We map each channel frequency to a corresponding value of $\lambda^2$. For each measurement we assume that $(Q, U, \sigma_{\rm Q}, \sigma_{\rm U})$ are recorded, where $Q$ corresponds to the real part of the polarization and $U$ corresponds to the imaginary component.

\subsection{Faraday Depth Spectra}
\label{sec:fdspectra}

%
%
We project the data into Faraday depth space following the numerical method outlined in \cite{2005A&A...441.1217B}. The resulting Faraday depth spectra are calculated using,
\begin{equation}
F(\phi) = K^{-1}\sum_{i = 1}^{N}{W_i P_i {\rm e}^{-2j (\lambda_i^2 - \lambda_0^2) \phi}},
\end{equation}
where $W_i$ is the weight for channel $i$ at a wavelength of $\lambda_i$, and $P_i$ is the complex polarization measured in that channel, see Equation~\ref{eq:pol}. These are the \emph{dirty} Faraday depth spectra, i.e. they are convolved with the RMTF, which itself is calculated using,
\begin{equation}
R(\phi) = K^{-1}\sum_{i = 1}^{N}{W_i {\rm e}^{-2j (\lambda_i^2 - \lambda_0^2) \phi}}.
\end{equation}
In each case, $K$, is the sum of the weights,
\begin{equation}
K = \sum_{i = 1}^{N}{W_i},
\end{equation}
where the weights are defined as the reciprocal of the variance in each channel, and $\lambda_0$ is a reference wavelength used to derotate the polarization data to a common zero-point defined as
\begin{equation}
\lambda_0^2 = K^{-1}\sum_{i = 1}^{N}{W_i \lambda_i^2}.
\end{equation}

\begin{figure*}
\centerline{\includegraphics[width=0.9\textwidth]{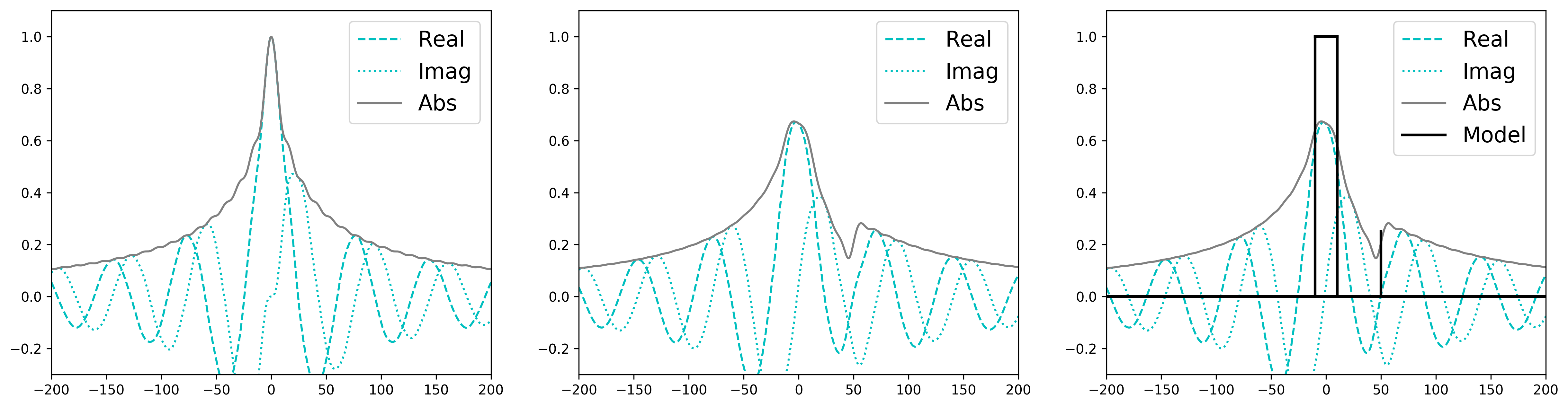}}
\centerline{\includegraphics[width=0.9\textwidth]{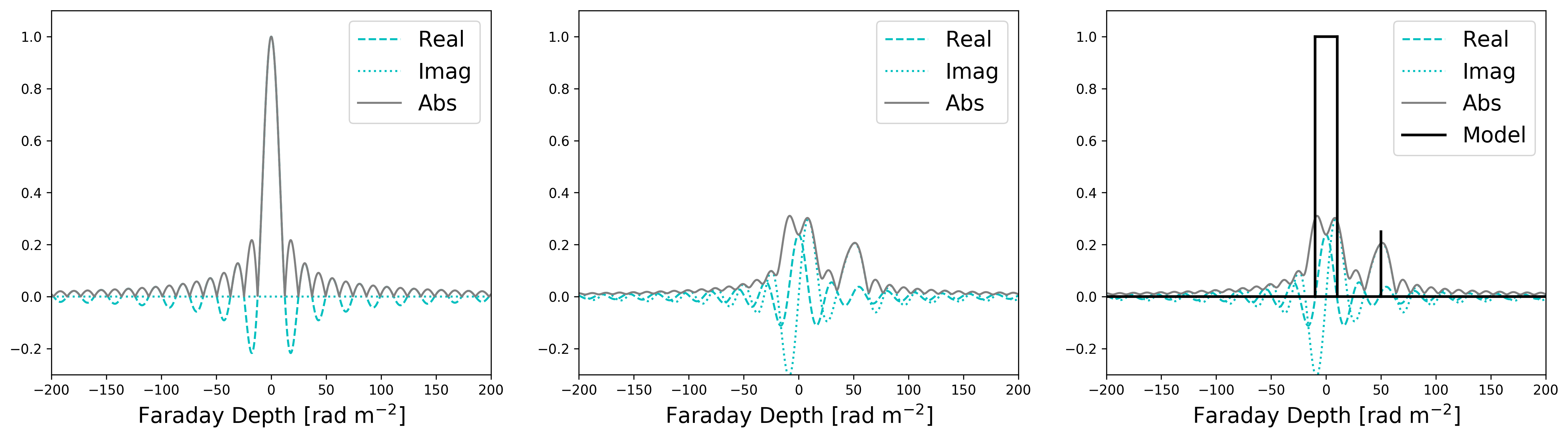}}
\caption{\label{fig:scen2samp} Scenario 2: RMTF (left), Faraday depth spectrum (centre) and Faraday depth spectrum overlaid with input model (right) for the case of (i) data evenly sampled in frequency (top) and (ii) data evenly sampled in wavelength-squared (bottom). Data are noiseless in both cases.}
\end{figure*}

The recovered Faraday depth spectra for Scenario~2 are shown in Figure~\ref{fig:scen2samp}, which illustrates the difference in the Faraday depth spectra when polarization data are sampled uniformly in frequency compared to being sampled uniformly in $\lambda^2$-space. From the figure, two key differences can be observed. Firstly, by comparing the FWHM of the RMTF in each case, it can be seen that the resolution in Faraday depth is improved by a factor of $\sim2$. This is important for resolving different Faraday components and complex structure. Secondly, it can be seen that the main lobe of the RMTF is clearly separated from the sidelobe structure in the case of uniform $\lambda^2$-sampling, and that the amplitude of the first sidelobe is $\sim3$ times smaller than that resulting from uniform sampling in frequency. As well as aiding visual inspection, this is important for approaches that incorporate deconvolution, as incorrect or imperfect identification of model component positions during deconvolution will result in residuals with amplitudes proportional to RMTF sidelobe levels. As a consequence of these factors, the Faraday thick structure from Scenario~2 in Figure~\ref{fig:scen2samp} is resolved in the resulting Faraday depth spectrum when using uniform $\lambda^2$-sampling, with the positions of the edges clearly delineated. \textcolor{black}{Additionally, we note that the sidelobe response of regular sampling in $\lambda^2$-space could be further improved by the application of a window function prior to Fourier transforming, and that such an application could be tailored to balance the contributions from sidelobe structure, thermal noise and resolution, as needed for an individual science case.}

In what follows, we use inverse variance weights for calculating Faraday depth spectra. For the optimised GP reconstructions, we use an approximate predictive posterior variance for the weights given by
\begin{equation}
\label{eq:gpvar}
\sigma_{\ast}^2 = C_{\ast} + \sigma_{\rm n}^2,
\end{equation}
\citep{3569} where $\sigma_{\rm n}^2$ is the amplitude of the white noise kernel in Equation~\ref{eq:whitekernel}. The white noise variance needs to be included because we are building a Faraday depth spectrum from the model estimate of the {\it noisy} complex polarization data.

\subsection{Parameter Estimation}
\label{sec:optimization}

To optimize the hyper-parameters of our GP kernel we use MCMC implemented via the {\sc emcee} package \citep{emcee} to explore the posterior probability distribution. We use 500 samples for a burn-in then take the maximum likelihood hyper-parameter values from the burn-in and perturb them by adding noise at a level of $10^{-5}$. We then use these perturbed values to start a production run of 5000 samples. Prior ranges for each of the hyper-parameters are shown in Table~\ref{tab:priors}. The lower limit on $\log P$ corresponds to a maximum Faraday depth of $\sim1200$\,rad\,m$^{-2}$.
\begin{table}
\caption{\label{tab:priors} Range of uniform priors used for each of the hyper-parameters in GP kernel, see Equation~\ref{eq:celkernel}.}
\centering
\begin{tabular}{|c|c|}
\hline
Hyper-parameter & Prior \\\hline
$\ln h$ & (-25,5)\\
$\ln c$ & (-25,5)\\
$\ln P$ & (-6,10)\\
$\ln \sigma$ & (-25,5)\\\hline
\end{tabular}
\end{table}

\subsection{Imputation of Missing Data}
\label{sec:missing}

RFI flagging creates gaps in the polarization data thereby reducing the $\lambda^{2}$-coverage in an irregular fashion and causing the sidelobe structure in the RMTF to become more prominent.
\begin{figure*}
\includegraphics[width=0.9\textwidth]{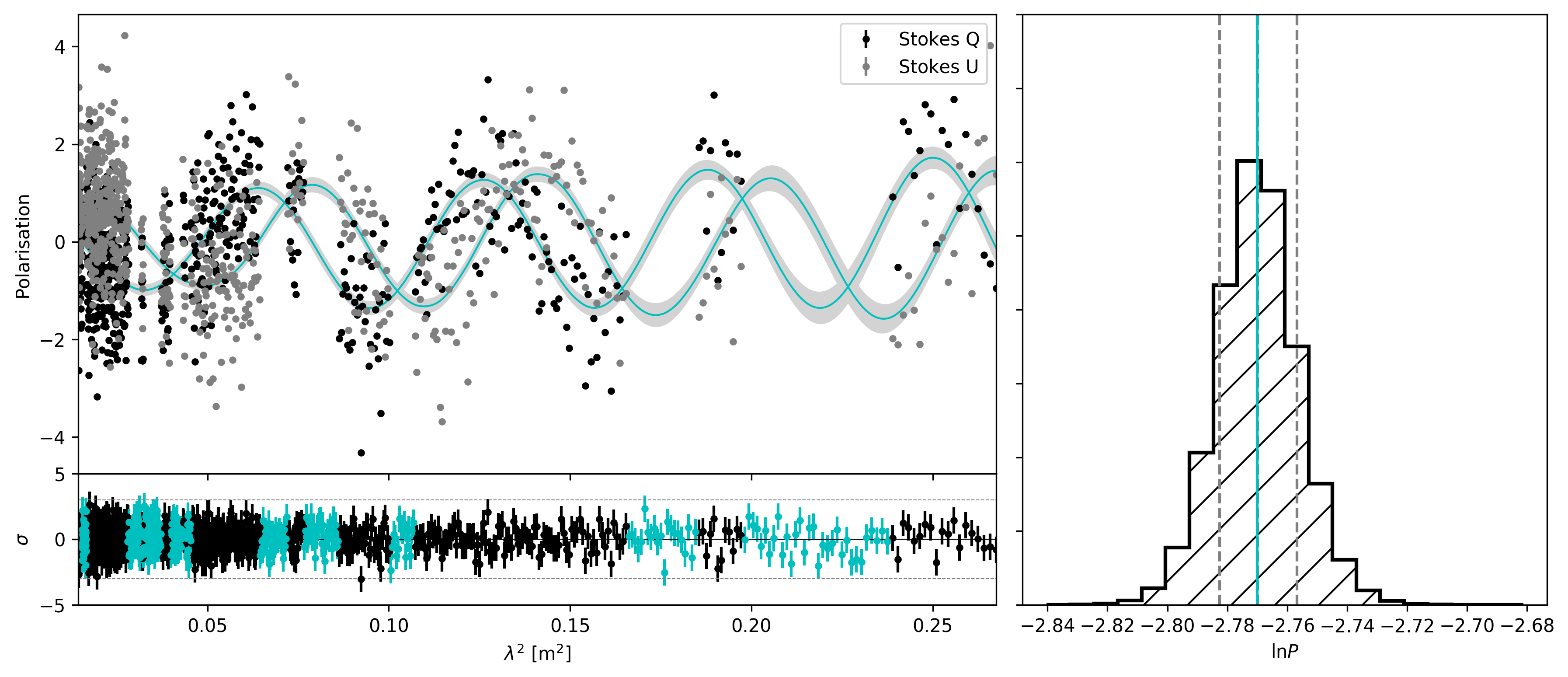}
\caption{\label{fig:scenario1} Left upper: Stokes parameters for Scenario 1 (see \S~\ref{sec:scenarios}) with SNR$_{\rm chan}=1$ and 30\% of the data flagged; black \& grey error bars), and maximum a posteriori GP model (Equation~\ref{eq:celkernel}; blue solid line) with one standard deviation posterior uncertainty (grey shaded area). Left lower: residuals between the posterior mean for Stokes Q and the data (black points) and residuals between the posterior mean and the missing/flagged Stokes Q data (blue points). The $\pm3\sigma$ limits are marked by grey dashed lines. Right: The inferred period of the process with the true period marked (vertical blue line) as well as the 16\%, 50\% and 84\% confidence intervals of the posterior probability distribution (grey dashed lines).}
\end{figure*}
To mimic the RFI flagging process, we remove random portions of data from the simulated data set. To do this, a random position in the data set is selected and a chunk comprised of a randomly selected number of entries is removed. This process is repeated until the desired percentage of data is removed. We simulate data sets with 20\%, 30\% and 40\% missing values. An example of this flagging using the Scenario~1 dataset is shown in Figure~\ref{fig:scenario1}, which shows the input data, flagged at a level of 30\%. These data are scattered by white noise, giving a signal-to-noise of one in each channel.

Following \cite{2018arXiv180604326S}, we define the SNR of the integrated polarization data as
\begin{equation}
{\rm SNR}_{\rm int} = \sqrt{N_{\rm chan}} \times \frac{P_0}{\sigma_{\nu}},
\end{equation}
where $\sigma_{\nu}$ is the standard deviation of the random Gaussian noise added to each frequency channel and $N_{\rm chan}$ is the number of frequency channels. $P_0$ is the polarized amplitude at the reference frequency, taken in this case to be 1.4\,GHz. Noise is added independently to both Stokes Q and Stokes U data.

Figure~\ref{fig:scenario1} also shows the posterior mean  from the GP model, optimized using only the unflagged data points, as well as the residual between the input data and the optimised GP model for both the unflagged and flagged data points. The right-hand panel of Figure~\ref{fig:scenario1} shows the posterior probability distribution for the hyper-parameter, $P$, which can be used to calculate the rotation measure as described later in this work, see Section~\ref{sec:rms}.

\begin{figure*}
\centerline{\includegraphics[width=0.95\textwidth]{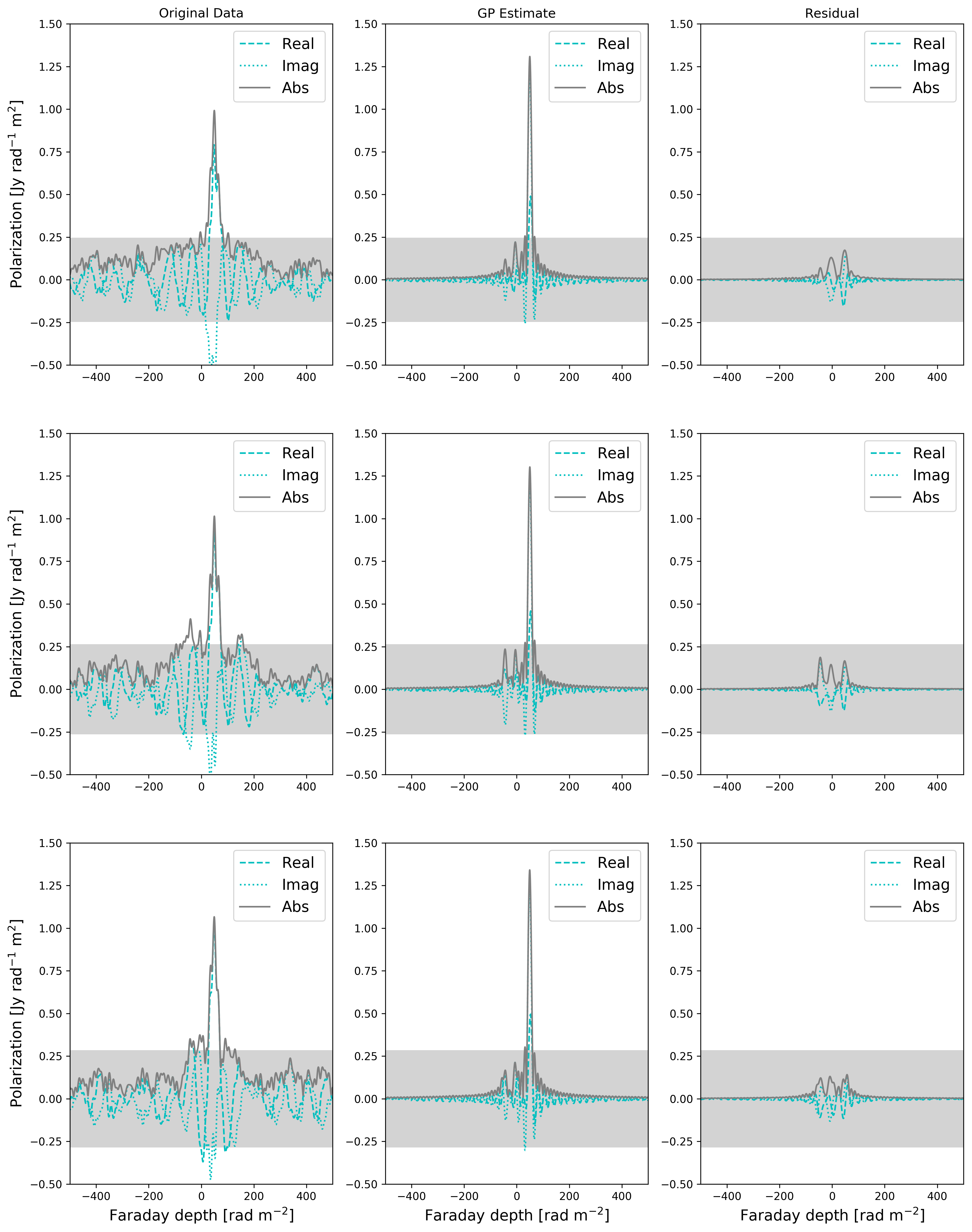}}
\caption{\label{fig:flagging1} Faraday depth spectra for Scenario~1 (Faraday thin) with 20\% (top), 30\% (middle) and 40\% (bottom) of channel data flagged and ${\rm SNR}_{\rm chan} = 1$. Left: Faraday depth spectrum from flagged data. Right: Faraday depth spectra from GPM reconstructed data as described in \S~\ref{sec:missing}. The $\pm$5\,$\sigma_{\phi}$ limits are shown as a light grey shaded area.}
\end{figure*}

The effect of the flagging process on the Faraday depth spectrum is shown for Scenario~1 in Figure~\ref{fig:flagging1}. In Figure~\ref{fig:flagging1}~(centre), data have been imputed at regular intervals across the full bandwidth of the simulation and the Faraday depth spectrum calculated using the posterior mean values, weighted by the posterior covariance at each point.

It is noticeable in Figure~\ref{fig:flagging1} that the spectra calculated from the optimized GP model appear significantly cleaner than those calculated from the original data. This is in part because the sidelobe structure due to the gaps in frequency coverage is reduced, but also because the posterior mean values from the model are not scattered by the thermal noise in the same way as the original data. Although this may be advantageous in enhancing some features of the spectra, it can also potentially lead to over-interpretation of low-level features. We caution that structures present at amplitudes below $5-8\,\sigma_{\phi}$ should not be considered to represent true astrophysical components but are more likely to arise from structure in the noise. For uniform spectral noise, i.e. noise per frequency channel, the value of $\sigma_{\phi}$ can be calculated by taking into account the number of independent measurements in complex polarization, which is $N_{\rm chan}'$, where $N_{\rm chan}'$ is the number of unflagged frequency channels in the data set. By analogy to the quantity $P(\lambda^2)$, which has $\sigma_P = \sigma$ when $\sigma_Q = \sigma_U = \sigma$ \citep{2005A&A...441.1217B}, the amplitude of the Faraday depth spectrum will have noise $\sigma_{\phi} = \sigma/\sqrt{N_{\rm chan}'}$. The threshold indicated in Figure~\ref{fig:flagging1} is shown at $5\sigma_{\phi}$.


The right-hand column of Figure~\ref{fig:flagging1} shows the  Faraday depth spectrum reconstructed from the difference between the GP estimate and the noiseless model in Scenario~1. We note that this difference is not the same as the residual that would be obtained from deconvolution of Faraday depth reconstruction of the GP estimate, which depends only on the data itself, but instead describes the degree of difference between the true model structure and the estimate fitted to the noisy data. Residuals in this structure below the level of the noise threshold are consistent with the model over-fitting to noise fluctuations in the Stokes~Q and U data from the perspective of the model. One notable feature of this difference in the case of low signal-to-noise is the appearance of a residual peak at $\phi = -50$\,rad\,m$^{-2}$ for Scenario~1. This is due to the fact that the GP model is not parametrically constrained to produce the same amplitude for both the Stokes~Q and U components, but rather the amplitude will follow the pattern of the noise in the data. For low signal to noise data, this can result in a local offsets between the amplitude of Q and U inferred by the GP estimate as a function of $\lambda^2$. From the perspective of the model this local offset removes flux from the true peak and moves it to the mirrored peak, resulting in the double structure seen in the residual Faraday depth spectrum. However, once again we note that from the perspective of the data, this is a consequence of the noise realisation, which is why it is reflected in the GP model estimate.

\begin{figure*}
\includegraphics[width=0.9\textwidth]{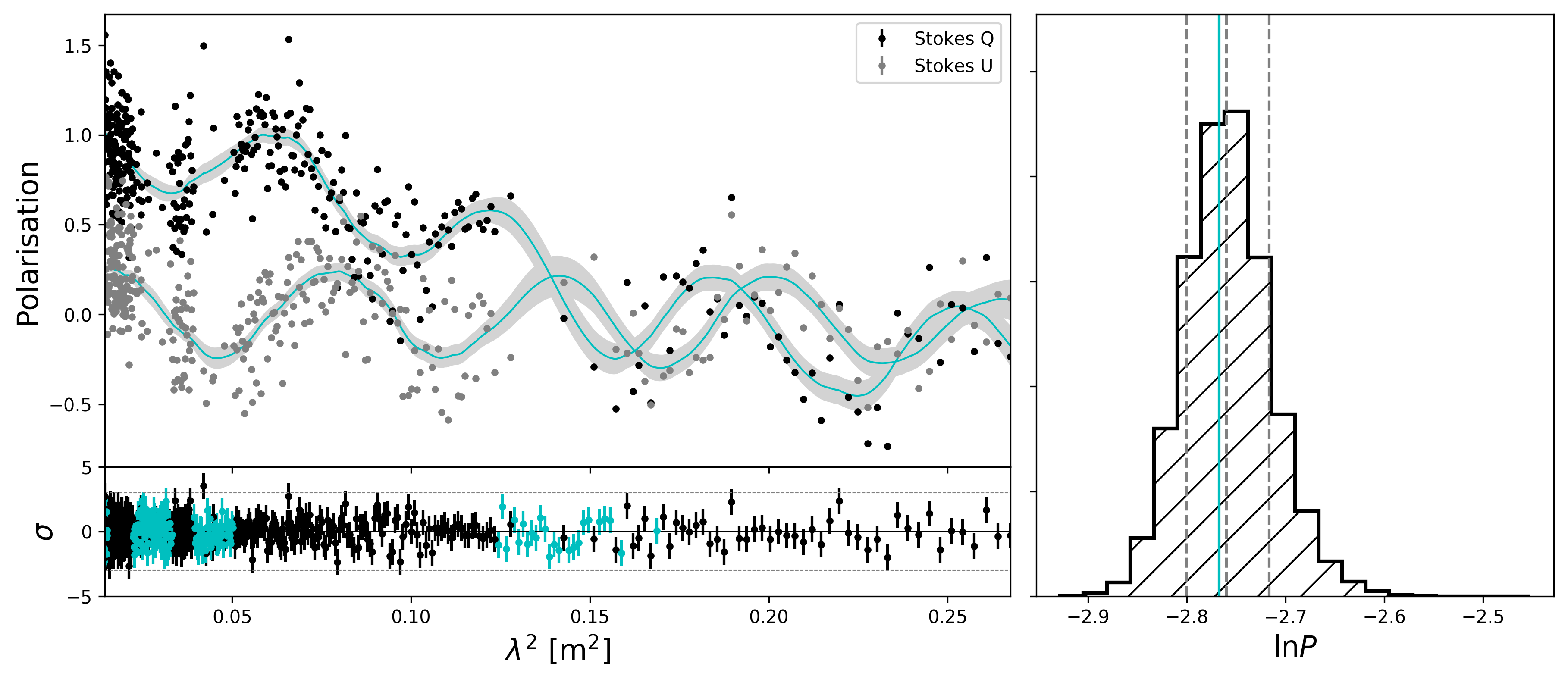}
\caption{\label{fig:scenario2} Left upper: Stokes parameters for Scenario 2 (see \S~\ref{sec:scenarios}; black \& grey error bars), and maximum a posteriori GP model (Equation~\ref{eq:celkernel}; blue solid line) with one standard deviation posterior uncertainty (grey shaded area). Left lower: residuals between the posterior mean for Stokes Q and the data (black points) and residuals between the posterior mean and the missing/flagged Stokes Q data (blue points). The $\pm3\sigma$ limits are marked by grey dashed lines. Right: The inferred period of the process with the true period marked (vertical blue line) as well as the 16\%, 50\% and 84\% confidence intervals of the posterior probability distribution (grey dashed lines).}
\end{figure*}

As illustrated by \cite{celerite}, it is often the case that valid inferences can be made using incorrect GP models. For physical processes whose underlying properties are not well understood, such as generalised Faraday spectra, this can significantly complicate the process of selecting a parametric model. Incorporating domain knowledge helps to make the problem of model selection and interpretation of parameters less difficult. We demonstrate this by using the GP kernel from Scenario~1 to impute missing data for simulations of Scenario~2. An example of this is shown in Figure~\ref{fig:scenario2}, which shows a simulated dataset where 20\% of the data points have been removed to mimic RFI flagging. In spite of not being an exact model, the optimized GP is able to reconstruct the missing data within the measurement uncertainties. It can also be seen that the optimized model is able to recover the rotation measure of the radio galaxy with high accuracy.

\begin{figure*}
\centerline{\includegraphics[width=0.95\textwidth]{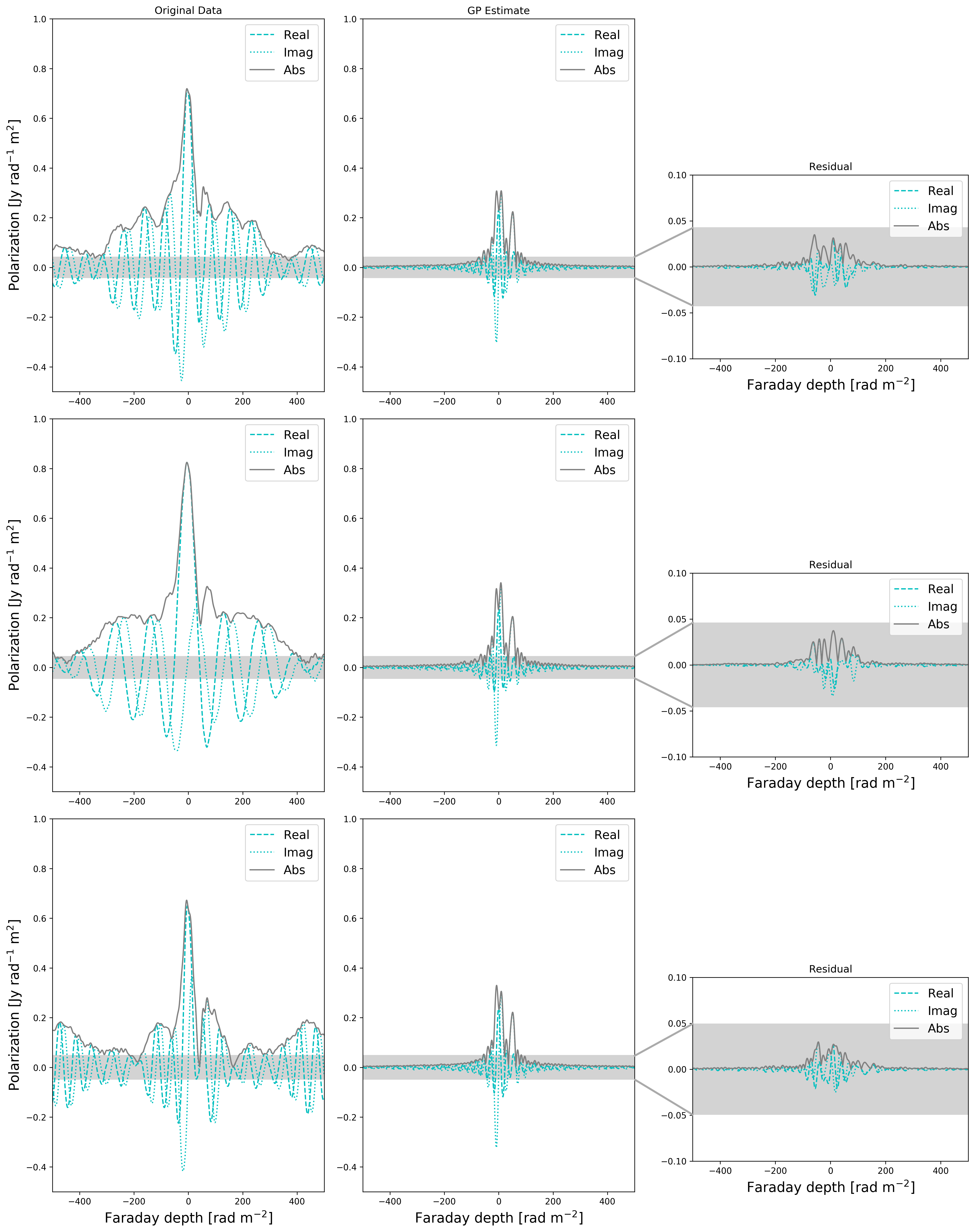}}
\caption{\label{fig:flagging2} Faraday depth spectra for Scenario~2 (Faraday complex) with 20\% (top), 30\% (middle) and 40\% (bottom) of channel data flagged and ${\rm SNR}_{\rm chan} = 5$ at a reference frequency of 1.4\,GHz. Left: Faraday depth spectrum from flagged data. Centre: Faraday depth spectra reconstructed from the GP estimate as described in \S~\ref{sec:missing}. Right: Faraday depth residuals formed by subtracting the noiseless model from the GP estimate. The $\pm$5\,$\sigma_{\phi}$ limits are shown as a light grey shaded area.}
\end{figure*}

In this case, recovering the Faraday depth of the thin component is not as strong an indicator of performance as in Scenario~1. Instead we compare the posterior mean of the optimized GP to the input model without noise and the residuals for this comparison are shown in the lower left hand panel of Figure~\ref{fig:scenario2}. Figure~\ref{fig:flagging2} shows the Faraday depth spectrum reconstructed from the optimized GP for Scenario~2 in the presence of different levels of RFI flagging. These should be compared to the noiseless model spectrum shown in Figure~\ref{fig:scen2samp}.

\subsection{Multiple Sources along the line of sight}
\label{sec:multiple}

\begin{figure*}
\includegraphics[width=0.88\textwidth]{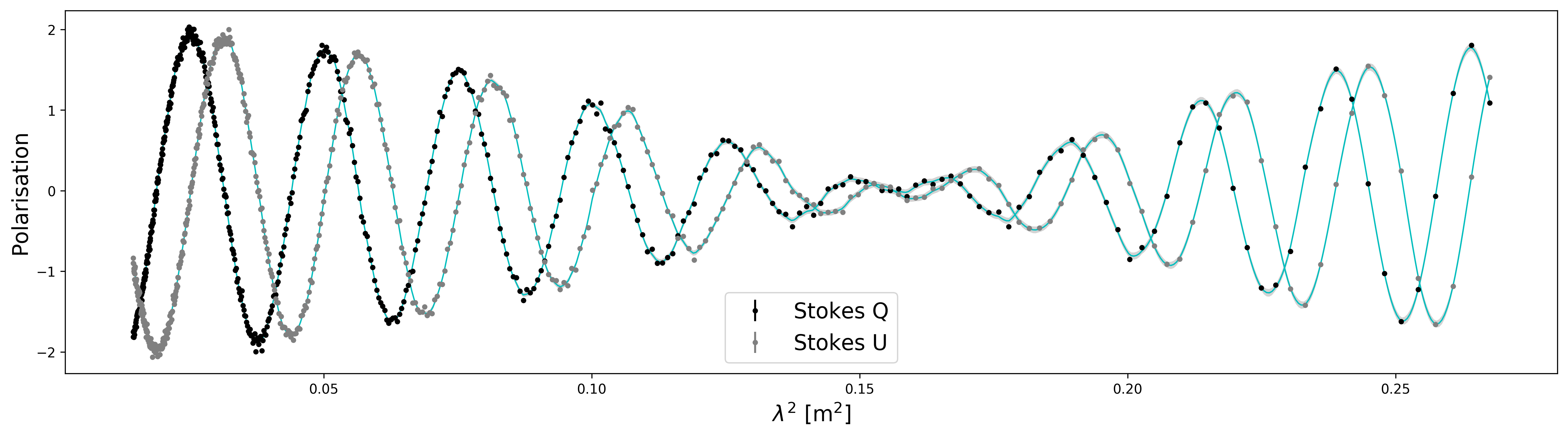}
\includegraphics[width=0.9\textwidth]{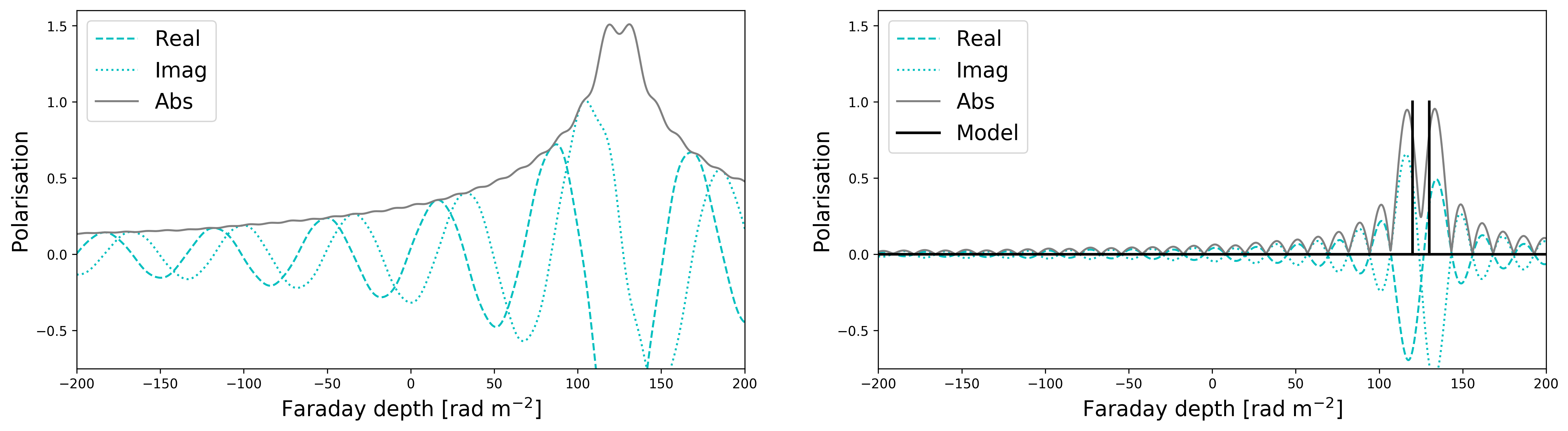}
\caption{\label{fig:double} Upper: Stokes parameters for line of sight comprised of two Faraday thin sources with SNR$_{\rm chan}=20$ (black \& grey error bars), and maximum a posteriori GP model (Equation~\ref{eq:celkernel}; blue solid line) with one standard deviation posterior uncertainty (grey shaded area). Left lower: Faraday depth spectrum calculated from the noisy input data sampled at regular intervals in frequency. Right lower: Faraday depth spectrum calculated from the optimised GPM sampled at regular intervals in $\lambda^2$-space with the model components marked as solid black lines.}
\end{figure*}

As described in \cite{Sun_2015}, it is expected that many lines of sight will have more than one Faraday thin component. Such Faraday geometries are also able to be recovered by the kernel in Equation~\ref{eq:celkernel}. In this case the hyper-parameter $P$ will represent the $|RM|$ of the component with the highest rotation. An example of where this can occur is the super-position of radio galaxy jets or lobes oriented along a line of sight unresolved by the beam of a telescope. Such geometries result in double Faraday components, typically with similar Faraday depths as the integrated rotation differs only by the degree contributed in the local medium between the jets whilst experiencing a common degree of rotation from the comparatively larger path between the galaxy and the observer. An example of such a situation is shown in Figure~\ref{fig:double}. We illustrate the GP model in this figure at high signal to noise, SNR$_{\rm chan} = 20$, to demonstrate that the same GP kernel from Equation~\ref{eq:celkernel} can accurately represent the complex polarization behaviour of multiple sources without requiring additional kernel components. The Faraday depth spectra in the lower panels of Figure~\ref{fig:double} show the recovered Faraday depth spectrum from the noisy input data (left) and from the optimised GP reconstruction (right). It can be observed that the double structure is more apparent in the GP reconstruction. We note that the separation of the peaks in Faraday depth is wider than the separation of components in the model. This is not an artifact of the GP reconstruction, but due to the position of the first null in the RMTF.

 \subsection{Rotation Measure Extraction}
\label{sec:rms}

Under certain circumstances it is possible to use the optimised hyper-parameters to recover the rotation measure of a source. In principle, for Faraday thin sources the absolute value of the rotation measure can be recovered from the periodicity using,
\begin{equation}
|{\rm RM}| = \frac{\pi}{P}.
\end{equation}
Due to the stationary nature of the covariance kernels, it is not possible to recover the sign of the rotation measure from the hyper-parameters of the covariance matrix; however, it can be recovered by comparing the correlation co-efficient, $c_{\rm corr}$, between (e.g.) the Stokes Q data and the Stokes U data shifted by $\pm\pi/2$ radians, $U^{\pm}$:
\begin{equation}
\label{eq:corr}
c_{\rm corr}^{\pm} = \langle Q,U^{\pm} \rangle.
\end{equation}

\begin{figure*}
\centerline{\includegraphics[width=0.9\textwidth]{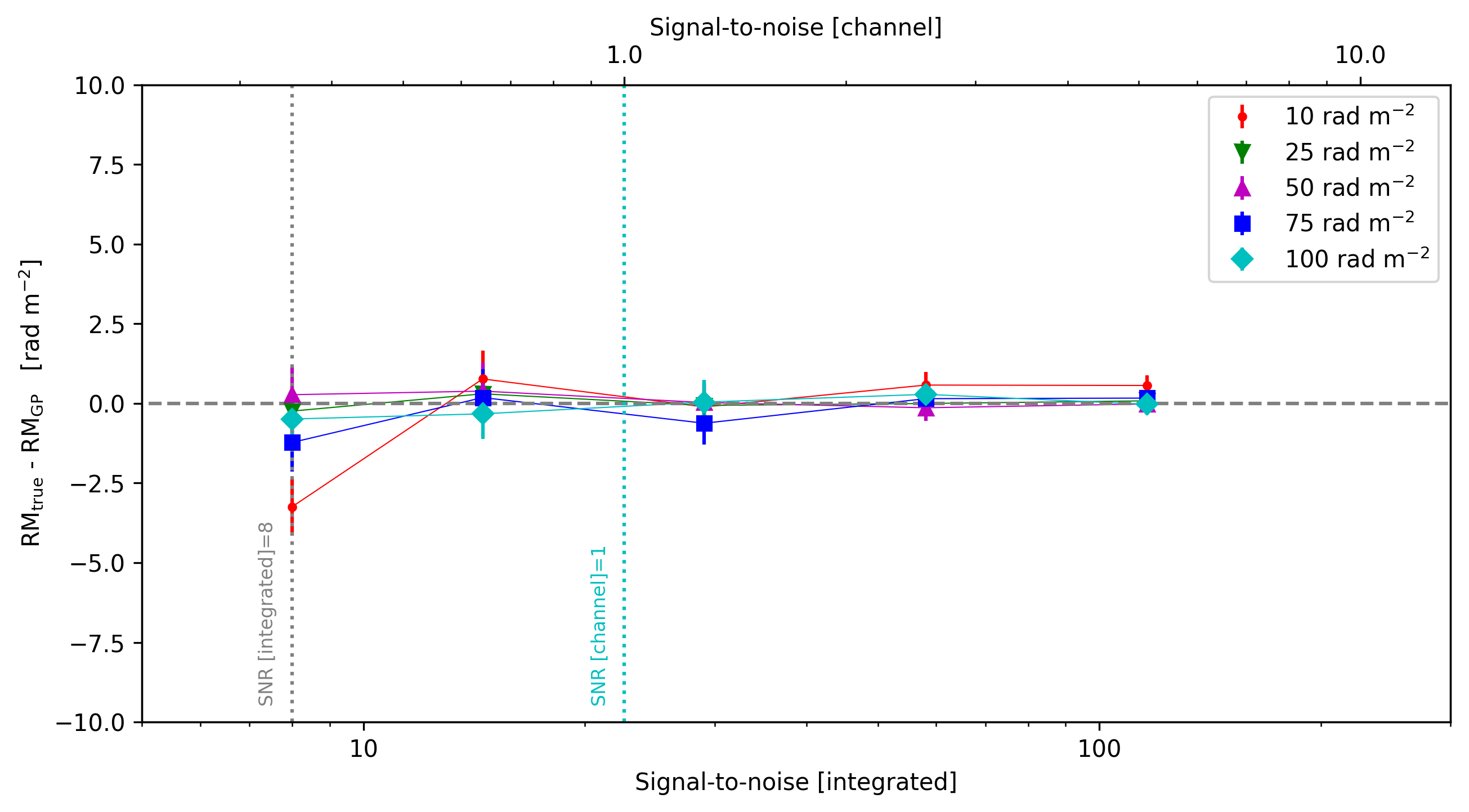}}
\caption{\label{fig:rm_snr} Recovered rotation measures (RM) as a function of the signal-to-noise (SNR) in the integrated polarization data for Scenario~1 are shown for a range of different $\phi_{\rm gal}$ values. The difference between the expectation of the RM from the GP and the input to the Scenario~1 model are shown in each case with uncertainties calculated using the 68\% confidence interval on the optimised value of $\log P$, which results in asymmetric uncertainties on the RM values. The point at which the signal-to-noise in an individual frequency channel is equal to one is shown as a vertical dashed line (blue). The distribution of equivalent expectation values for ten different noise realisations are shown as light grey lines.}
\end{figure*}

Using Scenario One, we evaluate the accuracy of RM recovery for the Faraday thin source as a function of the signal-to-noise (SNR) of the integrated polarization data, ${\rm SNR}_{\rm int}$. It can be seen in Figure~\ref{fig:rm_snr} that for Scenario~1, the rotation measure is recovered with $< 1\,\sigma$ deviation from the true value for even low values of ${\rm SNR}_{\rm int}<10$ where the signal-to-noise on an individual frequency channel, ${\rm SNR}_{\rm chan}\ll 1$. A minimum value of ${\rm SNR}_{\rm int}=8$ is typically used to identify sources in total polarization images, see e.g. \cite{george_stil_keller_2012}, and this value is marked in Figure~\ref{fig:rm_snr}. It can be seen that the optimized RM value is in good agreement with both the expectation value and the true value of the RM above this limit.

However, we note that when using this method to obtain an RM there is a lower limit on the absolute value of the rotation measure that can be reliably recovered. This limit is dependent on the both the frequency coverage of the observational data and the signal to noise ratio. At very low values of Faraday depth the degree of rotation in the complex polarization signal becomes smaller and causes a degeneracy between hyper-parameter $c$ in Equation~\ref{eq:celkernel} and the periodicity, $P$. This is exacerbated for data at higher frequencies where the rotation is more poorly sampled. This lack of rotation also causes the method for assigning the direction of the rotation measure described in Equation~\ref{eq:corr} to be less effective, as there are not guaranteed to be $\pm\pi/2$ radians of rotation present in the observed data. As this lower limit, $\phi_{\rm lim}$, is a strong function of frequency coverage, it is possible to determine it a priori for a particular telescope such that
\begin{equation}
\label{eq:rmlim}
    \phi_{\rm lim} = \frac{\pi}{4\Delta \lambda^2},
\end{equation}
where $\Delta \lambda^2$ is the coverage of the observational data in $\lambda^2$-space and the given expression corresponds to that coverage being equivalent to $\pi/2$\,radians for $\phi_{\rm lim}$.  A illustrated in Figure~\ref{fig:rm_snr}, RM values above this limit are recovered reliably for SNR$_{\rm int}\ge 8$.

We emphasise that this limit is only applicable to the calculation of the rotation measure from the hyper-parameters and does not affect the improvement in the Faraday depth spectrum from GPM re-sampling or imputation, as shown in Figures~\ref{fig:scen2samp}, where structure at low Faraday depths is recovered equivalently well.

\section{Performance considerations for surveys}
\label{sec:joedata}

We now extend the application to simulated data to a large data set, such as might be recovered from a MeerKAT observation of a survey field. We use a simulated data set containing Stokes~Q and Stokes~U data for 10,000 Faraday thin sources. Observational parameters are matched to those of the MeerKAT telescope, one of the precursor arrays of the Square Kilometre Array (SKA) \citep{5136190, 2009IEEEP..97.1522J}. The MeerKAT consists of 64 13.5\,m parabolic dishes sited in the Northern Cape province of South Africa \citep{Justin,Taylor_2017}. 


A population of 10,000 Faraday thin sources was created with a Stokes~I flux density at 1.4\,GHz taken from a uniform distribution with values between 10\,$\mu$Jy and  30\,mJy, and an RM taken from a uniform distribution between -100 and +100\,rad\,m$^{-2}$. Stokes~Q and U values were generated for \textcolor{black}{320 channels spaced uniformly in frequency} from 0.88\,GHz to 1.68\,GHz. The intrinsic angle of polarisation for each source, $\chi_0$, was sampled randomly from a uniform distribution $[-\pi, +\pi)$.
For each frequency channel, noise was added to the Stokes~Q and U values using a normal distribution with a mean of zero and a variance of $180\mu$\,Jy\,beam$^{-1}$, equivalent to an integrated noise level of $\sim 10\mu$Jy\,beam$^{-1}$. These specifications are representative of the observational parameters for the MeerKAT MIGHTEE survey \citep{Taylor_2017}.


The frequency-dependence of the flux density was modelled such that
\begin{equation}
\label{FreqDependent}
S_{\nu} = S_{\rm 1.4\,GHz}\left(\frac{\nu}{\rm 1.4\,GHz} \right)^{-\alpha},
\end{equation}
where $S_\nu$ is the flux density at each frequency, $\nu$, $S_{1.4\rm{GHz}}$ is the flux density at 1.4 $\rm{GHz}$, and $\alpha$ is the spectral index, which is taken to be the canonical value of $\alpha$ = 0.7 for synchrotron radiation.

The intensity of the polarised signal $P$ is given by
\begin{equation}
\label{eq:PolarisedSignal}
P = \sqrt{Q^2 + U^2} = I*\Pi,
\end{equation}
where $Q$ and $U$ are the Stokes parameters for linear polarisation, $I$ is the total intensity Stokes parameter, and $\Pi$ is the polarisation fraction. To simulate values for $\Pi$, the following relationship was used,
\begin{equation}
\label{eq:StilEquation}
\log \Pi = -(0.051\pm0.004) \log S_{\rm 1.4\,GHz} + (0.388\pm0.007)
\end{equation}
\citep{Stil_2014}, where $\Pi$ is the median percentage polarisation fraction and $S_{\rm 1.4\,GHz}$ is the 1.4\,GHz flux density in mJy.
%
%
%

We randomly select 500 sources from this population with an SNR$_{\rm int}\ge8$ and $|RM|>\phi_{\rm lim}$ and use an optimised GP with the kernel specified in Equation~\ref{eq:celkernel} to recover the rotation measure for each object. We use a prior ranges as specified in Table~\ref{tab:priors}.
Uncertainties on the recovered RMs are calculated from the 68\% confidence interval on the optimised value of $\log P$, which results in asymmetric uncertainties on the RM values. For 500 randomly selected samples under these constraints all of the recovered RMs lie within 5$\sigma$ of the true value and $>99\%$ within 3$\sigma$. The average run time per object is 43\,seconds on a Macbook Pro with a 3.5\,GHz Intel Core i7 processor.
%


\section{Discussion}
\label{sec:disc}

\subsection{Using GP reconstructed Faraday depth spectra}
\label{sec:usage}

Examples of Faraday depth spectra reconstructed from the optimized GP for Scenario~1 and Scenario~2 are shown in Figures~\ref{fig:flagging1}~\&~\ref{fig:flagging2}. The reconstructed Faraday depth spectra in each case (right hand column of both figures) represent the structure in the quasi-periodic component of the covariance kernel only. Although the white noise component is taken into account during the optimization, it is not appropriate to scatter the predicted values by this distribution as the white component represents the measurement noise on the original data only. As mentioned already in \S~\ref{sec:missing}, in order to avoid over-interpretation of the reconstructed Faraday depth spectrum, \textcolor{black}{the significance of features below the equivalent detection threshold in Faraday depth space should be carefully considered.}

If the optimized GPM was used to infer the posterior mean only at the positions of the input data, the model could be subtracted from the data at those positions to estimate a residual, as shown in the lower panel of Figures~\ref{fig:scenario1}~\&~\ref{fig:scenario2}. This residual could then be transformed separately to the model and combined with the model reconstruction in a manner similar to that employed by the final step of the Cotton-Schwab CLEAN algorithm in synthesis imaging. We note that a kernel-based GPM approach to two-dimensional image synthesis and deconvolution in radio astronomy is likely to be prohibitive in its computational complexity, as the fast separable kernels employed here are only appropriate for one-dimensional data. However, kernel-free methods such as those described by \cite{arras2020} can achieve similar scaling.

\subsubsection{Recovery of Faraday thick structure}
\label{sec:thick}

As can be seen in Figure~\ref{fig:flagging2}, using an optimized GP to reconstruct the Faraday depth spectrum in the case where Faraday thick structure is present can reveal these structures without the need for assuming any specific parametric model of Faraday depth a priori. Although it is not possible to recover the parameterised geometry of the thick structure from the hyper-parameters of the GP in this case, the shape is far more clearly delineated in the GP-reconstructed Faraday depth spectrum than in the spectrum calculated from the original data. This remains the case even in the presence of high degrees of RFI excision.

In principle, GPs might also be used to infer the complex polarization signal at smaller values of $\lambda^2$ than are present in the original dataset, which could be beneficial for recovering extended structure in Faraday depth space. When used to infer the behaviour of a signal outside the range of measurements, the posterior covariance of a GP will increase in response to the stationary nature of the covariance not finding any additional measurements to anchor it. This increase in covariance can be used to down-weight predicted values of the posterior mean, relative to those known with more confidence. The behaviour of the inferred posterior mean will tend towards the mean function and in the general case where the mean function is assumed to be zero-valued, this is equivalent to the same logical sampling as in the original dataset; however, for spectra where Faraday thick structure are present, which have non-zero means at low-$\lambda^2$ this may artificially truncate structure. This was found to be the case here for the kernel in Equation~\ref{eq:celkernel} when used to extend $\lambda^2$-space.

It is also important to note that structures which exist only on scales larger than that set by the measured $\lambda^2_{\rm min}$ will {\it not} be recovered by a GP model used for inferring data at positions outside the range of the original measurements. Hence this form of structure enhancement for Faraday thick structures can only be used to improve the recovery of structures which exist on scales that are already sampled in part by the original measurements. Any larger structures will not appear in the Faraday depth spectrum of the GP reconstruction, an effect analogous to flux loss in synthesis imaging for radio interferometry.

\subsubsection{Zero-valued Faraday depth signals}
\label{sec:leakage}

In the scenarios considered here, we have assumed a zero-valued mean function. In the absence of instrumental leakage, this assumption is valid for modelling sources with non-zero Faraday rotation; however, there are also two scenarios where a non-zero offset in both Stokes~Q and Stokes~U can occur, resulting in a zero-valued Faraday depth signal. The first of these is the case of un-rotated radio polarization signals, the second is the presence of residual instrumental leakage. The inclusion of such a constant offset as a free parameter (separately for Q and U) in the optimization process is straightforward to implement and requires only very minor changes to be made in the calculation of the posterior mean. However, for some telescopes with wide bandwidths, polarization leakage can be frequency-dependent and consequently modelling it as a constant offset will not be sufficient. In this case there is potential for confusion between the leakage term and covariate behaviour in the complex polarization; as can be seen in Figure~\ref{fig:scenario2}, smoothly varying systematic changes in amplitude can be modelled perfectly well using stationary kernels.

\subsection{Faraday Challenge}
\label{sec:challenges}

%

%
In order to evaluate the performance of the GP approach against other methods in the literature we use the 17 models from the Faraday challenge of \cite{Sun_2015}. These models include examples of single Faraday thin objects, multiple (two) Faraday thin objects, and combinations of  Faraday thick and Faraday thin objects. While the GP approach can only determine RM values in the case of single Faraday thin models, we can use the model evaluation metrics to compare the predicted Faraday depth spectra in all cases. For this purpose we use the kernel from Equation~10 for all test data sets with priors as listed in Table~\ref{tab:priors}. All evaluations are made on models with SNR$_{\rm int}$=32, as in \cite{Sun_2015}.

The reduced $\chi^2$ for parametric models described in \cite{Sun_2015} is roughly equivalent to the standardized mean squared error (SMSE). For an optimised GP, the SMSE is defined as
\begin{equation}
\label{eq:smse}
SMSE = \frac{1}{N}\left[ \frac{(P - P_{\ast})^2}{\sigma_{\rm n}^2}  \right],
\end{equation}
where $P_{\ast}$ is the posterior mean  of complex polarization from the optimized GP, $P$ is the input complex polarization, and $\sigma_{\rm n}^2$ is the variance of the white noise on the input data. In order to account for the independence of the white noise between Stokes~Q and U we set $N=600$. The difference between the SMSE and the $\chi^2_{\rm r}$ used by \cite{Sun_2015} is the normalisation, which instead uses $N-K$ in the denominator where $K$ is the number of parameters. Although the GP used here is not strictly parametric, we calculate this quantity using $K=4$ to represent the number of hyper-parameters in our model.

Since the GP optimisation method produces a full predictive distribution at the position of each test sample, we also calculate the mean standardised log likelihood as a measure of performance \citep{3569}. The negative log-likelihood for the GP is given by,
\begin{equation}
\label{eq:logl}
-\log L^{\ast} = \frac{1}{2}\log(2\pi \sigma_{\ast}) + \frac{(P - P_{\ast})^2}{2\sigma_{\ast}^2},
\end{equation}
where $P$ are the input polarization data, $P_{\ast}$ are the GP reconstructed polarization data, and $\sigma_{\ast}^2$ is the variance of the GP reconstruction, defined in Equation~\ref{eq:gpvar}.
We standardise this loss by subtracting the equivalent loss as calculated using a model that predicts using a simple Gaussian with the mean and variance of the input data. The individual performance metrics for each model in the Faraday challenge are listed in Table~\ref{tab:challenge}.

\newcommand\Tstrut{\rule{0pt}{2.6ex}}       
\newcommand\Bstrut{\rule[-1.1ex]{0pt}{0pt}} 
\begin{table*}

\caption{Faraday challenge models. This table lists the input parameters for the 17 Faraday challenge models in Columns 1 to 9, and the evaluation metrics for the predicted complex Stokes data from the optimised GP in Columns 10 to 12. See Section~\ref{sec:challenges} for details. \label{tab:challenge} }

\begin{tabular}{@{\extracolsep{2pt}}|l|c|c|c|c|c|c|c|c|c|c|c|c|c|@{}}
\hline
 & \multicolumn{9}{c}{Input Parameters} &  \multicolumn{4}{c}{Evaluation Metrics} \\
  \cline{2-10}  \cline{11-14} \Tstrut
Model  & $p_1$ & $\phi_1$ & $\chi_1$ & $p_2$ & $\phi_2$ & $\chi_2$ & $p_0$ & $\phi_{\rm c}$ & $\phi_{\rm s}$ & $\chi^2_r$ & SMSE & MSLL &$\langle \phi-$SMSE$\rangle$  \\
 & $\%$ & rad\,m$^{-2}$ & deg & $\%$ & rad\,m$^{-2}$ & deg & $\%$ & rad\,m$^{-2}$ & rad\,m$^{-2}$ & & & & \\\hline
01 & 100.0 & 500.10 & 40 & $-$ & $-$ & $-$ & $-$ & $-$ & $-$ & 0.88 & 0.87 & -1.46 & 1.62 \\
02 & 100.0 & 49.38 & 60 & $-$ & $-$ & $-$ & $-$ & $-$ & $-$ & 1.00 & 0.99 & -0.94 & 1.31 \\
03 & 100.0 & 4.96 & 60 &  $-$ &  $-$ &  $-$ & $-$ & $-$ & $-$   & 1.06 & 1.05 & -0.47 & 1.01 \\
04 & 25.0 & $-37.84$ & 0 & 16.70 & 103.18 & $-36$ & $-$ & $-$ & $-$  & 1.03 & 1.03 & -0.99 & 1.33 \\
05 & 25.0 & $-37.84$ & $-40$ & 24.00 & 5.05 & $-40$ & $-$ & $-$ & $-$  & 0.96 & 0.95 & -0.71 & 1.18 \\
06 & 25.0 & $-37.84$ & $-40$ & 9.00 & 5.05 & $-40$ & $-$ & $-$ & $-$  & 1.05 & 1.04 & -0.80 & 1.16 \\
07 & 25.0 & $-44.55$ & 0 & 16.70 & 37.50 & 72 & $-$ & $-$ & $-$  & 1.08 & 1.07 & -0.51 & 1.28 \\
08 & 25.0 & 232.56 & 40 & 9.00 & 192.70 & 40 & $-$ & $-$ & $-$ & 1.04 & 1.03 & -1.27 & 1.30 \\
09 & 25.0 & $-37.83$ & $-40$ & 16.50 & 5.05 & 140 & $-$ & $-$ & $-$ & 0.91 & 0.90 & -0.72 & 1.05 \\
10 & 25.0 & $-37.84$ & 0 & 9.00 & 103.00 & $-36$ & $-$ & $-$ & $-$ & 1.05 & 1.05 & -0.90 & 1.27 \\
11 & 25.0 & 149.50 & $40$ & 23.75 & 163.50 & $-68$ & $-$ & $-$ & $-$ & 1.10 & 1.09 & -1.34 & 1.33 \\
12 & 25.0 & $-232.56$ & 0 & 9.00 & $-50.10$ & 72 & $-$ & $-$ & $-$  & 1.03 & 1.02 & -1.47 & 1.41 \\
13 & 25.0 & $-44.55$ & 0 & 24.00 & 37.54 & 72 & $-$ & $-$ & $-$   & 0.99 & 0.98 & -0.58 & 1.17 \\
14 &  $-$ & $-$ & $-$ & $-$ & $-$ & $-$ & 1.9 & $-136.98$ & 50  & 0.87 & 0.87 & -1.41 & 1.59 \\
15 & 1.8 & $-240.22$ & $-36$ & $-$ & $-$ & $-$ & 1.9 & $-250.17$ & 50 & 1.07 & 1.06 & -1.35 & 1.16 \\
16 & $-$ & $-$ & $-$ & $-$ & $-$ & $-$ & 1.9 & $-136.98$ & 25 & 0.99 & 0.98 & -1.53 & 1.25 \\
17 & 1.8 & $-240.00$ & $-36$ & $-$ & $-$ & $-$ & 1.9 & $-250.17$ & 25 & 1.02 & 1.01 & -1.41 & 1.20 \\\hline
\end{tabular}
\end{table*}

\begin{figure}
    \centering
    \includegraphics[width=0.45\textwidth]{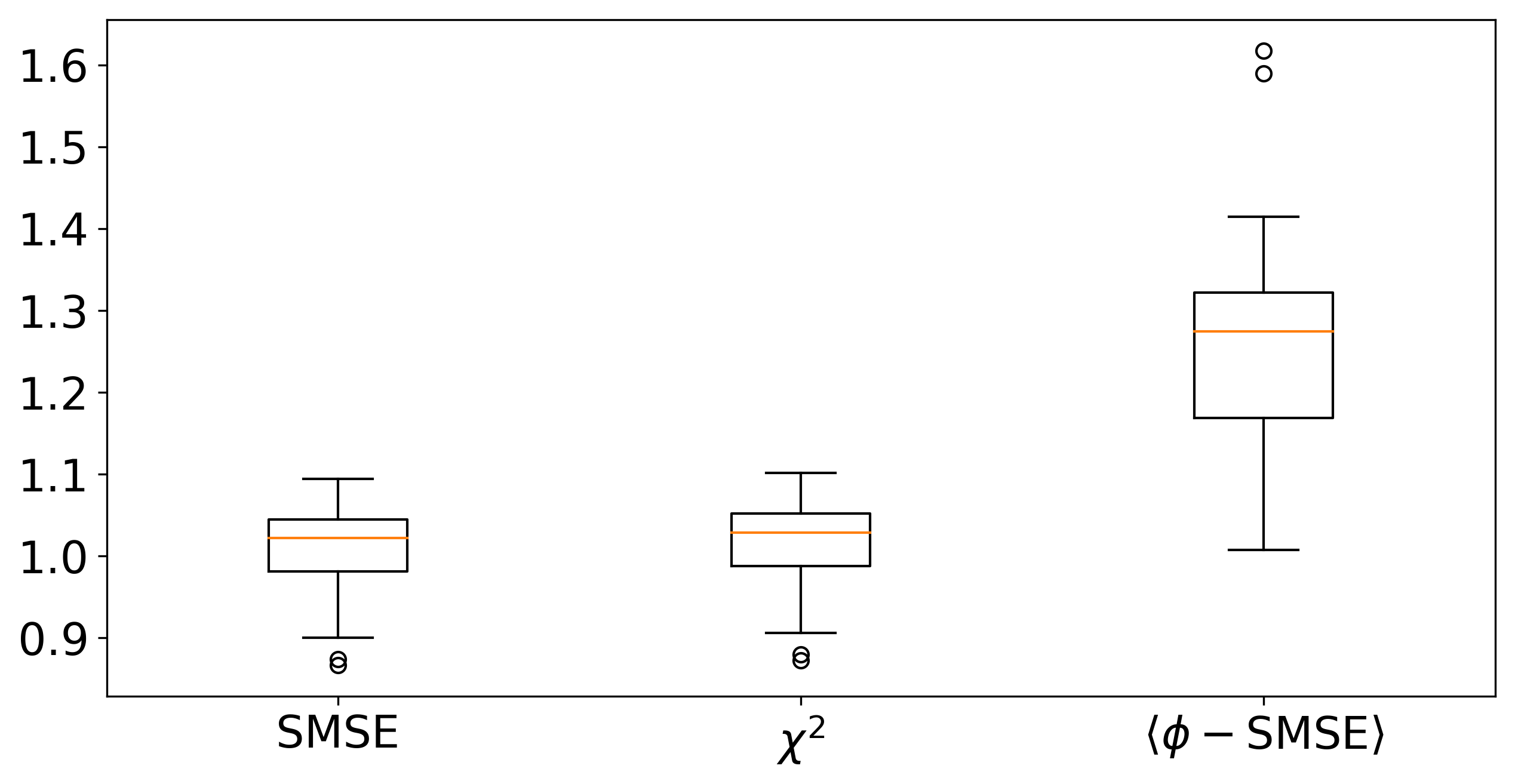}
    \caption{Box-and-whisker plots for figures of merit over all 17 Faraday challenge models. The boxes show the first, the second (median), and the third quartile, and the ends of the whiskers show the edges of the distribution. Points deemed to be outliers are shown as open circles.}
    \label{fig:sunmetrics}
\end{figure}

The median of the reduced chi-squared statistic recovered from our model, $\chi^2_{\rm r} = 0.99$, is in good agreement with the expected optimum value of $1.00\pm0.02$ from \cite{Sun_2015}. Results for both the SMSE and $\chi_{\rm r}^2$ across all 17 test models, evaluated using SNR=32 as in \cite{Sun_2015}, are shown in Figure~\ref{fig:sunmetrics}. By comparison to Figure~3 in that work it can be seen that the GP $\chi_{\rm r}^2$ has a median value comparable to that of the three best performing methods in that work, each of which use QU-fitting approaches. Furthermore, it can also be seen that the distribution of $\chi_{\rm r}^2$ for the GP method is significantly narrower than those methods, each of which has a 68\% confidence interval that extends to values larger than $\chi_{\rm r}^2=1.1$.

%
%

In order to evaluate the resulting Faraday depth reconstructions we draw $T=100$ Monte Carlo realisations from the posterior distribution of the optimised GP for each model. We then construct a residual Faraday depth spectrum from the difference between the model realisation and the noiseless Faraday challenge model. To mitigate the effect of noise correlation in Faraday depth space, we down-sample these residual Faraday depth data by a factor
\begin{equation}
    f = \Delta \phi_{\rm FWHM} / \delta \phi,
\end{equation}
where $\Delta \phi_{\rm FWHM} \simeq 135$\,rad\,m$^{-2}$ is the FWHM of the RMTF for the Faraday challenge data and $\delta \phi$ is the pixel size in Faraday depth. We use the down-sampled residual Faraday depth spectra to calculate an SMSE of the form shown in Equation~\ref{eq:smse},  relative to a model of $0 + 0j$. The Faraday depth SMSE uses $\sigma_{\phi} = \sigma /\sqrt{N_{\rm chan}}$ as a normalisation, as described in  Section~\ref{sec:missing}, and $N = 2\times N_{\rm ds}$, where $N_{\rm ds}$ is the number of data points in each of the down-sampled complex parts of the residual Faraday depth spectrum. We average the resulting SMSE values across all realisations for a particular model to obtain the quantity $\langle \phi$-SMSE$\rangle$ and this is listed for each model in Table~\ref{tab:challenge} and the distribution visualised in Figure~\ref{fig:sunmetrics}.

By comparing the SMSE and $\langle \phi$-SMSE$\rangle$ values for each model it can be seen that the goodness of fit for the GP model estimates is higher in the Stokes~Q and U data space than it is in the Faraday depth signal space. This is to be expected as the model itself is optimised against the noisy Stokes data. Figure~\ref{fig:sunmetrics} also indicates models that are deemed to be outliers based on the interquartile range of the data. These two outliers are Models~1 and 14, which are low outliers for the data (QU) space SMSE distribution and high outliers for the signal ($\phi$) space $\langle \phi$-SMSE$\rangle$ distribution, indicating that the GP model estimates are over-fitting the noisy QU data in these cases, resulting in larger Faraday depth residuals.

\begin{figure*}
\includegraphics[width=0.95\textwidth]{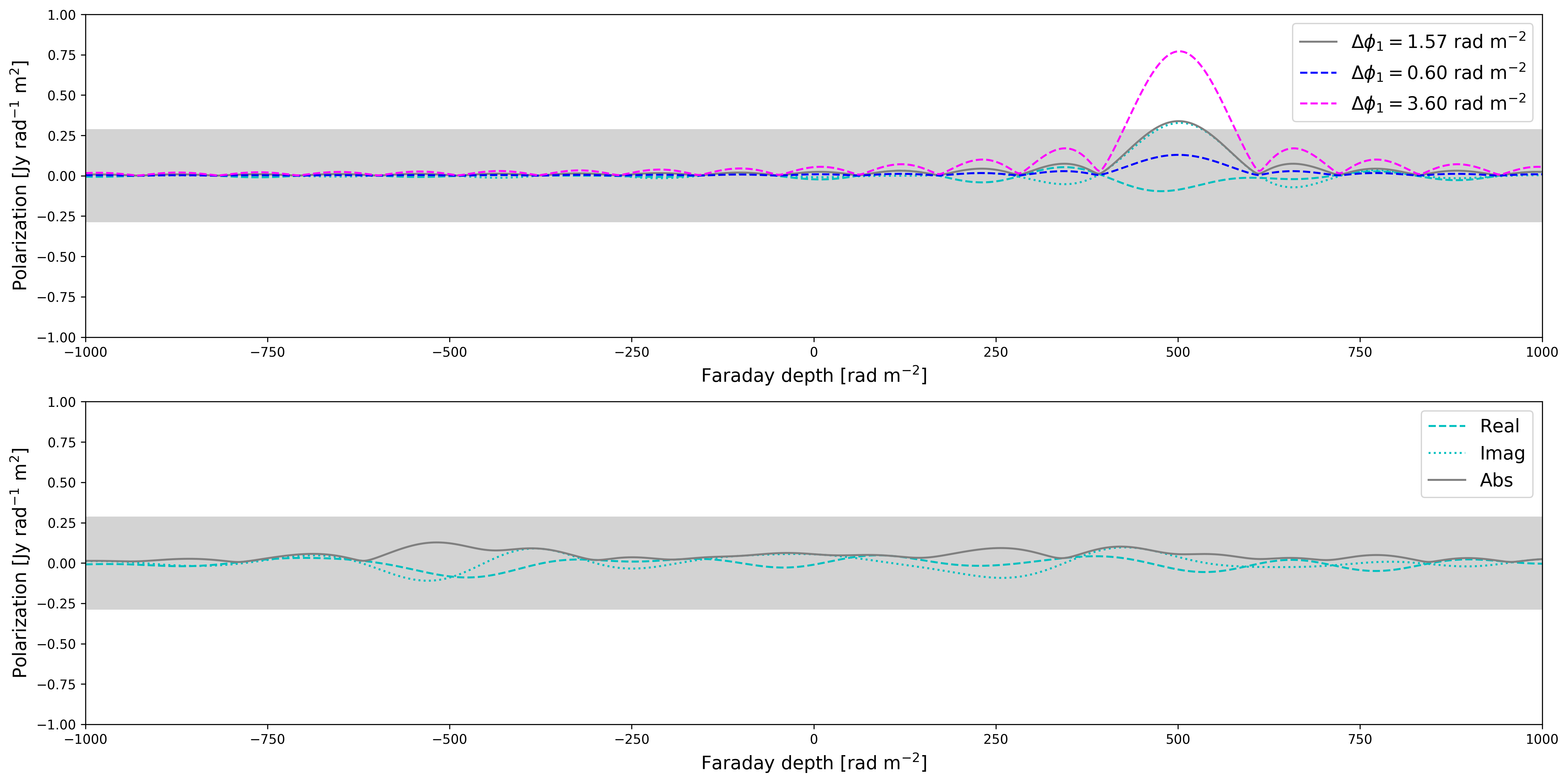}
\caption{Upper: Residual Faraday depth spectra for Model~1 of the Faraday challenge, \textcolor{black}{constructed using the transform of} the difference between the noiseless model and a model where the fitted Faraday depth, $\phi_1$, differs from the input model by (i) an average deviation of $\Delta \phi_1 = 1.57$\,rad\,m$^{-2}$ (grey line indicates absolute polarization, with real and imaginary components shown in cyan), (ii) a minimum deviation of $\Delta \phi_1 = 0.60$\,rad\,m$^{-2}$ (blue dashed line), and (iii) a maximum deviation of $\Delta \phi_1 = 3.60$\,rad\,m$^{-2}$ (magenta dashed line). Lower: Residual Faraday depth spectrum for Model~1 of the Faraday challenge, \textcolor{black}{constructed using the transform of the difference between the noiseless model and} the posterior mean of the optimised GP model. The $\pm$5\,$\sigma_{\phi}$ limits appropriate to SNR$_{\rm int}=32$ for the Faraday challenge are shown as a light grey shaded area. \label{fig:residuals}}
\end{figure*}

The calculation of $\langle \phi$-SMSE$\rangle$ performed here is possible due to the nature of the GP model, which provides a full predictive posterior distribution for each estimate. Whilst equivalent values are not available for the other methods profiled in the Faraday challenge of \cite{Sun_2015}, we are able to visualise the equivalent residual Faraday depth spectra. For example, for Model~1, the average difference between the Faraday depth, $\phi_1$ (see Table~\ref{tab:challenge}), of the noiseless model and the fitted value of this parameter across all of the methods in the \cite{Sun_2015} paper is $\Delta \phi_1 = 1.57$\,rad\,m$^{-2}$ (see Table~3 in that work). In Figure~\ref{fig:residuals} we show the residual Faraday depth spectra for the difference between the true noiseless model and a model with a value of $\phi_1$ that is incorrect by this amount. We also include residuals for $\Delta \phi_1 = 0.60$\,rad\,m$^{-2}$ and $\Delta \phi_1 = 3.60$\,rad\,m$^{-2}$, which were the smallest and largest differences returned by the methods evaluated against Model~1 in the \cite{Sun_2015} comparison. It can be seen from this figure that the amplitude of the residuals from the GP model are equivalent to those of the best fitting model from the \cite{Sun_2015} Faraday challenge in this case; however, it should also be noted that the nature of these residuals is different in form to those resulting from an inexact parametric model.

\subsection{Alternative kernels}
\label{sec:infocrit}

For a given regression problem, there are an infinite number of models that can be used. As such, it is usually not feasible to test all possible models. In this work so far we have used the GP kernel defined in Equation~\ref{eq:celkernel}; here we consider the use of alternative kernels. We denote our original kernel as as ``3-HP" and we compare it to (i) the quasi-periodic kernel used by \cite{celerite} to model stellar variability, and (ii) a hybrid kernel including a second squared-exponential term.

In \cite{celerite}, the period of rotation of a star was determined by extracting the optimised $P$ hyper-parameter, analogous to our use of $P$ to estimate the rotation measure of a Faraday thin source. The stellar variability kernel from \cite{celerite} is also built using the {\tt celerite} library and has the form,
\begin{equation}
\label{eq:4hpkernel}
k(x_i,x_j)=\frac{B}{1+C}\,{\rm e}^{-A|x_i - x_j|}\left[\cos\left(\frac{2\pi|x_i - x_j|}{P}\right) + (1+C)\right],
\end{equation}
where $A$, $B$, $C$ and $P$ are hyper-parameters. We denote this kernel as ``4-HP".

The hybrid kernel, denoted here as ``5-HP", has the form,
\begin{equation}
\label{eq:5hpkernel}
k(x_i,x_j)=h_1\,{\rm e}^{-c_1|x_i - x_j|}\cos\left(\frac{2\pi|x_i - x_j|}{P}\right) + h_2\,{\rm e}^{-c_2|x_i - x_j|},
\end{equation}
where $h_1$, $c_1$, $P$, $h_2$ and $c_2$ are hyper-parameters. Both the 4-HP and 5-HP kernels have an extra term which is intended to model smoothly varying non-periodic behaviour such as that which might arise from Faraday thick components along the line of sight.

Although different methods of evaluating competing models have been developed, the computational complexity involved in applying them limits their applicability. To overcome this limitation, easier-to-use approximate methods are commonly used to compare the performance of two or more models. These include the Bayesian Information Criterion (BIC), the Akaike Information Criterion (AIC) and the corrected Akaike Information Criterion (AICc). These information criteria are designed to evaluate the amount of information lost by a particular model. The model that loses the least information, i.e. has the lowest valued information criterion, is considered to be the preferred model.

\cite{burnham} suggest that the  BIC is  a powerful model selection tool when the `true' model is in the set of test models; however, in the case of empirical data where the true model is unlikely to be known a priori the AIC is considered to be a more appropriate criterion. Furthermore, the \emph{corrected} AIC places a stronger complexity penalty on models used with finite data series than the original AIC. In the case of small sample sizes it is considered to be more appropriate and accurate for finding the preferred model. However, \cite{kassrafferty} suggest that the AIC should only be used in situations where the prior distribution is similar to that of the likelihood and that when this isn't the case and the prior is significantly less informative than the likelihood then the BIC will favour simpler models than the AIC.

For each sample dataset we optimize the hyper-parameters for each of the three kernels using the method described in \S~\ref{sec:optimization} and use these values to compute the Bayesian Information Criteria \citep[BIC;][]{1978AnSta...6..461S},
\begin{equation}
{\rm BIC} = -2 \log L^\ast + K \log N,
\end{equation}
the Akaike Information Criterion \citep[AIC;][]{AIC},
\begin{equation}
{\rm AIC} = 2 K - 2\log L^\ast,
\end{equation}
and the corrected AIC \citep[AICc;][]{AICc},
\begin{equation}
{\rm AICc} = {\rm AIC} + \frac{2K(K+1)}{N - K - 1},
\end{equation}
where $\log L^\ast$ is given by Equation~\ref{eq:logl}, $K$ is the number of hyper-parameters, and $N=600$ is the number of data points.

We calculate the median value and inter-quartile range for each of the three information criteria from ten different noise realisations with SNR$_{\rm int}$=32.
%
The goodness of fit statistics for the Faraday challenge models using the two alternative kernels are $\chi^2_{\rm r} = 0.98$ for the 4-HP model, and $\chi^2_{\rm r} = 0.97$ for the 5-HP model.

\begin{figure}
    \centering
    \includegraphics[width=0.45\textwidth]{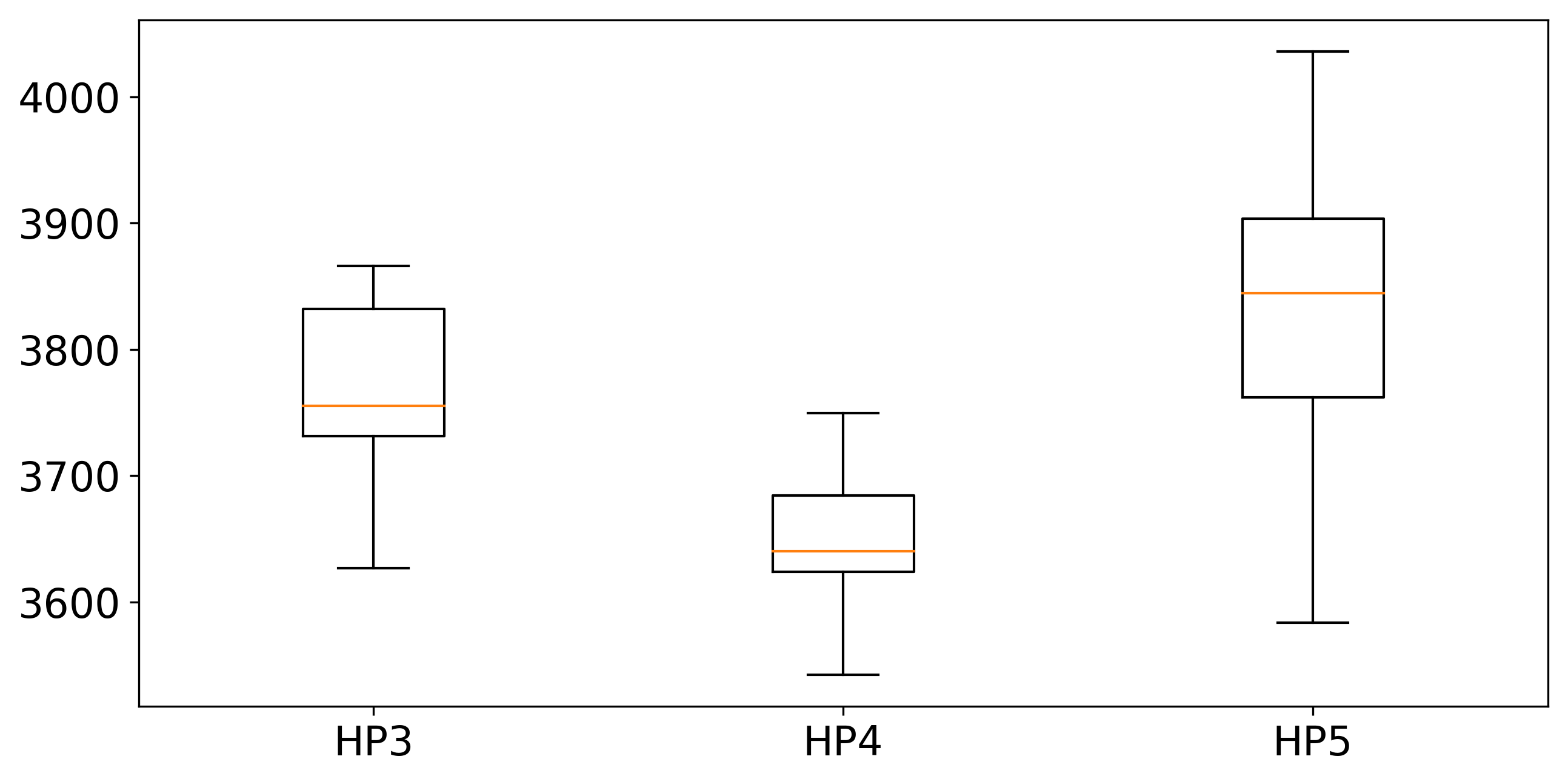}
    \caption{Box-and-whisker plots for BIC values derived from each of the 3 kernels using 10 noise realisations of Model 1 from \protect\cite{Sun_2015}. The boxes show the first, the second (median), and the third quartile, and the ends of the whiskers show the minimum and maximum values.}
    \label{fig:model1bic}
\end{figure}

The median values of the information criteria suggest a preference for the HP-4 kernel in the case of single Faraday thin component models, a preference for the HP-5 kernel in the case of double Faraday thin models, and a preference for the HP-3 kernel in the case of Faraday thick models. However, we note that these are modal preferences and no model type shows a unanimous preference for a particular kernel. Indeed, although ranking the three kernels using the median value of the information criteria for each data set may nominally show a preference for one kernel above others, the overlapping inter-quartile ranges for most models suggest that there is in fact no clear preference. The only model where one kernel is significantly preferred is Model 1, where the inter-quartile range of the HP-4 kernel is outside the inter-quartile range of both other kernels, see Figure~\ref{fig:model1bic}. We also see no difference in the median ranking across the three information criteria, which is perhaps unsurprising given that $N \gg K$ in all cases. Furthermore, as noted by \cite{biccomp}, information criteria comparisons are appropriate only for data sets with the same number of measurement samples, and we emphasise that the kernel comparison presented here is for the specifications of the test sets in the \cite{Sun_2015} Faraday challenge.

\section{Conclusions}
\label{sec:conclusions}

In this work we suggest that Gaussian Process Modelling (GPM) may be a flexible pre-processing approach for the improved analysis of Faraday depth spectra as it does not rely on an exact parameterised physical model, or set of models, to be known a priori and is therefore suitable for situations where some prior knowledge of the form of the signal is understood, but the exact line of sight structure is imperfectly known. This is often the case for Faraday depth analysis. With regard to this, we have shown that even simple GPM covariance kernels can be used to impute data reliably for complex lines of sight.

We have demonstrated the applicability of GPM for improved recovery of structure in Faraday depth spectra. We have shown that these improvements include the reduction of sidelobe contamination in Faraday depth spectra through imputation of missing data when RFI flagging causes significant losses, even in the case where significant portions of the original data set are lost. We have also illustrated the benefits of using GPM to re-sample complex polarization data onto regular spacings in $\lambda^2$-space as an alternative to currently used re-gridding methods. We have shown that these applications in combination can provide an improved representation of Faraday thick structure in Faraday depth spectra.

We have shown that under certain circumstances the hyper-parameters of the periodic covariance kernel can be used to recover rotation measure values for Faraday thin sources and that this hyper-parameter can also be physically meaningful in a more limited way for complex lines of sight. We have presented a method to calculate the range of RM over which this method is applicable for different telescopes and/or data sets and we have demonstrated that within this range RM recovery from GP hyper-parameters is robust to typical detection thresholds.

We have discussed limitations in the use of GPM reconstructed Faraday depth spectra, highlighted situations in which GPM is not an appropriate solution and presented a quantitative method for determining a threshold above which Faraday structure can be safely considered significant when reconstructed using an optimised GP fitted to noisy data.

We have evaluated the GP reconstruction proposed here against other methods in the literature and have shown that it has a data space performance equivalent to other QU-fitting methods and furthermore that the variance in results from the GP approach is smaller in the Stokes data space than seen from those methods. \textcolor{black}{Using the same test data we demonstrate that even complex Faraday structure can be recovered using a comparatively simple covariance
kernel.}

\section*{Acknowledgements}

We thank the anonymous referee for their careful reading of the manuscript and their helpful comments, which improved the content of this paper. The authors also thank Torsten En{\ss}lin for useful comments on an earlier draft of this paper. This work makes extensive use of the {\tt celerite} library\footnote{\url{https://celerite.readthedocs.io}}. SWN acknowledges support from the UK Newton Fund via the Science \& Technology Facilities Council (STFC) through the Development in Africa with Radio Astronomy (DARA) Big Data project ST/R001898/1. AMS gratefully acknowledges support from an Alan Turing Institute AI fellowship EP/V030302/1. JH acknowledges support from the UK Science \& Technology Facilities Council (STFC). MC acknowledges support from ANID PFCHA/DOCTORADO BECAS CHILE/2018-72190574.

\section*{Data Availability}

All code from this work is publicly available on Github at the following  address: {\tt https://github.com/as595/GPM-for-Faraday-Depth-Spectra }




\bibliographystyle{mnras}
\bibliography{gpm_paper} 

\begin{thebibliography}{}
\makeatletter
\relax
\def\mn@urlcharsother{\let\do\@makeother \do\$\do\&\do\#\do\^\do\_\do\%\do\~}
\def\mn@doi{\begingroup\mn@urlcharsother \@ifnextchar [ {\mn@doi@}
  {\mn@doi@[]}}
\def\mn@doi@[#1]#2{\def\@tempa{#1}\ifx\@tempa\@empty \href
  {http://dx.doi.org/#2} {doi:#2}\else \href {http://dx.doi.org/#2} {#1}\fi
  \endgroup}
\def\mn@eprint#1#2{\mn@eprint@#1:#2::\@nil}
\def\mn@eprint@arXiv#1{\href {http://arxiv.org/abs/#1} {{\tt arXiv:#1}}}
\def\mn@eprint@dblp#1{\href {http://dblp.uni-trier.de/rec/bibtex/#1.xml}
  {dblp:#1}}
\def\mn@eprint@#1:#2:#3:#4\@nil{\def\@tempa {#1}\def\@tempb {#2}\def\@tempc
  {#3}\ifx \@tempc \@empty \let \@tempc \@tempb \let \@tempb \@tempa \fi \ifx
  \@tempb \@empty \def\@tempb {arXiv}\fi \@ifundefined
  {mn@eprint@\@tempb}{\@tempb:\@tempc}{\expandafter \expandafter \csname
  mn@eprint@\@tempb\endcsname \expandafter{\@tempc}}}

\bibitem[\protect\citeauthoryear{{Aigrain}, {Parviainen}  \& {Pope}}{{Aigrain}
  et~al.}{2016}]{2016MNRAS.459.2408A}
{Aigrain} S.,  {Parviainen} H.,   {Pope} B.~J.~S.,  2016, \mn@doi [\mnras]
  {10.1093/mnras/stw706}, \href
  {https://ui.adsabs.harvard.edu/abs/2016MNRAS.459.2408A} {459, 2408}

\bibitem[\protect\citeauthoryear{{Akaike}}{{Akaike}}{1974}]{AIC}
{Akaike} H.,  1974, \mn@doi [IEEE Transactions on Automatic Control]
  {10.1109/TAC.1974.1100705}, 19, 716

\bibitem[\protect\citeauthoryear{{Angus}, {Morton}, {Aigrain}, {Foreman-Mackey}
   \& {Rajpaul}}{{Angus} et~al.}{2018}]{2018MNRAS.474.2094A}
{Angus} R.,  {Morton} T.,  {Aigrain} S.,  {Foreman-Mackey} D.,   {Rajpaul} V.,
  2018, \mn@doi [\mnras] {10.1093/mnras/stx2109}, \href
  {https://ui.adsabs.harvard.edu/abs/2018MNRAS.474.2094A} {474, 2094}

\bibitem[\protect\citeauthoryear{{Arras}, {Perley}, {Bester}, {Leike},
  {Smirnov}, {Westermann}  \& {En{\ss}lin}}{{Arras} et~al.}{2020}]{arras2020}
{Arras} P.,  {Perley} R.~A.,  {Bester} H.~L.,  {Leike} R.,  {Smirnov} O.,
  {Westermann} R.,   {En{\ss}lin} T.~A.,  2020, arXiv e-prints, \href
  {https://ui.adsabs.harvard.edu/abs/2020arXiv200811435A} {p. arXiv:2008.11435}

\bibitem[\protect\citeauthoryear{{Barclay}, {Endl}, {Huber}, {Foreman-Mackey},
  {Cochran}, {MacQueen}, {Rowe}  \& {Quintana}}{{Barclay}
  et~al.}{2015}]{2015ApJ...800...46B}
{Barclay} T.,  {Endl} M.,  {Huber} D.,  {Foreman-Mackey} D.,  {Cochran} W.~D.,
  {MacQueen} P.~J.,  {Rowe} J.~F.,   {Quintana} E.~V.,  2015, \mn@doi [\apj]
  {10.1088/0004-637X/800/1/46}, \href
  {https://ui.adsabs.harvard.edu/abs/2015ApJ...800...46B} {800, 46}

\bibitem[\protect\citeauthoryear{Bollen, Harden, Ray  \& Zavisca}{Bollen
  et~al.}{2014}]{biccomp}
Bollen K.~A.,  Harden J.~J.,  Ray S.,   Zavisca J.,  2014, \mn@doi [Structural
  equation modeling : a multidisciplinary journal]
  {10.1080/10705511.2014.856691}, 21, 1

\bibitem[\protect\citeauthoryear{{Bond} \& {Efstathiou}}{{Bond} \&
  {Efstathiou}}{1987}]{1987MNRAS.226..655B}
{Bond} J.~R.,  {Efstathiou} G.,  1987, \mn@doi [\mnras]
  {10.1093/mnras/226.3.655}, \href
  {https://ui.adsabs.harvard.edu/abs/1987MNRAS.226..655B} {226, 655}

\bibitem[\protect\citeauthoryear{Brandenburg \& Stepanov}{Brandenburg \&
  Stepanov}{2014}]{Brandenburg_2014}
Brandenburg A.,  Stepanov R.,  2014, \mn@doi [The Astrophysical Journal]
  {10.1088/0004-637x/786/2/91}, 786, 91

\bibitem[\protect\citeauthoryear{{Brentjens} \& {de Bruyn}}{{Brentjens} \& {de
  Bruyn}}{2005}]{2005A&A...441.1217B}
{Brentjens} M.~A.,  {de Bruyn} A.~G.,  2005, \mn@doi [\aap]
  {10.1051/0004-6361:20052990}, \href
  {https://ui.adsabs.harvard.edu/abs/2005A&A...441.1217B} {441, 1217}

\bibitem[\protect\citeauthoryear{{Burn}}{{Burn}}{1966}]{1966MNRAS.133...67B}
{Burn} B.~J.,  1966, \mn@doi [\mnras] {10.1093/mnras/133.1.67}, \href
  {https://ui.adsabs.harvard.edu/abs/1966MNRAS.133...67B} {133, 67}

\bibitem[\protect\citeauthoryear{Burnham \& Anderson}{Burnham \&
  Anderson}{2004}]{burnham}
Burnham K.~P.,  Anderson D.~R.,  2004, \mn@doi [Sociological Methods \&
  Research] {10.1177/0049124104268644}, 33, 261

\bibitem[\protect\citeauthoryear{Czekala, Andrews, Mandel, Hogg  \&
  Green}{Czekala et~al.}{2015}]{Czekala_2015}
Czekala I.,  Andrews S.~M.,  Mandel K.~S.,  Hogg D.~W.,   Green G.~M.,  2015,
  \mn@doi [The Astrophysical Journal] {10.1088/0004-637x/812/2/128}, 812, 128

\bibitem[\protect\citeauthoryear{{Dewdney}, {Hall}, {Schilizzi}  \&
  {Lazio}}{{Dewdney} et~al.}{2009}]{5136190}
{Dewdney} P.~E.,  {Hall} P.~J.,  {Schilizzi} R.~T.,   {Lazio} T. J. L.~W.,
  2009, \mn@doi [Proceedings of the IEEE] {10.1109/JPROC.2009.2021005}, 97,
  1482

\bibitem[\protect\citeauthoryear{{Evans}, {Aigrain}, {Gibson}, {Barstow},
  {Amundsen}, {Tremblin}  \& {Mourier}}{{Evans}
  et~al.}{2015}]{2015MNRAS.451..680E}
{Evans} T.~M.,  {Aigrain} S.,  {Gibson} N.,  {Barstow} J.~K.,  {Amundsen}
  D.~S.,  {Tremblin} P.,   {Mourier} P.,  2015, \mn@doi [\mnras]
  {10.1093/mnras/stv910}, \href
  {https://ui.adsabs.harvard.edu/abs/2015MNRAS.451..680E} {451, 680}

\bibitem[\protect\citeauthoryear{Farnsworth, Rudnick  \& Brown}{Farnsworth
  et~al.}{2011}]{Farnsworth_2011}
Farnsworth D.,  Rudnick L.,   Brown S.,  2011, \mn@doi [The Astronomical
  Journal] {10.1088/0004-6256/141/6/191}, 141, 191

\bibitem[\protect\citeauthoryear{{Foreman-Mackey}, {Hogg}, {Lang}  \&
  {Goodman}}{{Foreman-Mackey} et~al.}{2013}]{emcee}
{Foreman-Mackey} D.,  {Hogg} D.~W.,  {Lang} D.,   {Goodman} J.,  2013, \mn@doi
  [\pasp] {10.1086/670067}, \href
  {https://ui.adsabs.harvard.edu/abs/2013PASP..125..306F} {125, 306}

\bibitem[\protect\citeauthoryear{{Foreman-Mackey}, {Agol}, {Angus}  \&
  {Ambikasaran}}{{Foreman-Mackey} et~al.}{2017}]{celerite}
{Foreman-Mackey} D.,  {Agol} E.,  {Angus} R.,   {Ambikasaran} S.,  2017, ArXiv

\bibitem[\protect\citeauthoryear{George, Stil  \& Keller}{George
  et~al.}{2012}]{george_stil_keller_2012}
George S.~J.,  Stil J.~M.,   Keller B.~W.,  2012, \mn@doi [Publications of the
  Astronomical Society of Australia] {10.1071/AS11027}, 29, 214–220

\bibitem[\protect\citeauthoryear{{Gibson}, {Aigrain}, {Roberts}, {Evans},
  {Osborne}  \& {Pont}}{{Gibson} et~al.}{2012}]{2012MNRAS.419.2683G}
{Gibson} N.~P.,  {Aigrain} S.,  {Roberts} S.,  {Evans} T.~M.,  {Osborne} M.,
  {Pont} F.,  2012, \mn@doi [\mnras] {10.1111/j.1365-2966.2011.19915.x}, \href
  {https://ui.adsabs.harvard.edu/abs/2012MNRAS.419.2683G} {419, 2683}

\bibitem[\protect\citeauthoryear{{Haywood} et~al.,}{{Haywood}
  et~al.}{2014}]{2014MNRAS.443.2517H}
{Haywood} R.~D.,  et~al., 2014, \mn@doi [\mnras] {10.1093/mnras/stu1320}, \href
  {https://ui.adsabs.harvard.edu/abs/2014MNRAS.443.2517H} {443, 2517}

\bibitem[\protect\citeauthoryear{{Heald}}{{Heald}}{2009}]{2009IAUS..259..591H}
{Heald} G.,  2009, in {Strassmeier} K.~G.,  {Kosovichev} A.~G.,   {Beckman}
  J.~E.,  eds,  IAU Symposium Vol. 259, Cosmic Magnetic Fields: From Planets,
  to Stars and Galaxies. pp 591--602, \mn@doi{10.1017/S1743921309031421}

\bibitem[\protect\citeauthoryear{Hurvich \& Tsai}{Hurvich \& Tsai}{1989}]{AICc}
Hurvich C.~M.,  Tsai C.-L.,  1989, \mn@doi [Biometrika]
  {10.1093/biomet/76.2.297}, 76, 297

\bibitem[\protect\citeauthoryear{{Ideguchi}, {Tashiro}, {Akahori}, {Takahashi}
  \& {Ryu}}{{Ideguchi} et~al.}{2014}]{2014ApJ...792...51I}
{Ideguchi} S.,  {Tashiro} Y.,  {Akahori} T.,  {Takahashi} K.,   {Ryu} D.,
  2014, \mn@doi [\apj] {10.1088/0004-637X/792/1/51}, \href
  {https://ui.adsabs.harvard.edu/abs/2014ApJ...792...51I} {792, 51}

\bibitem[\protect\citeauthoryear{{Jonas}}{{Jonas}}{2009}]{2009IEEEP..97.1522J}
{Jonas} J.~L.,  2009, \mn@doi [IEEE Proceedings] {10.1109/JPROC.2009.2020713},
  \href {https://ui.adsabs.harvard.edu/abs/2009IEEEP..97.1522J} {97, 1522}

\bibitem[\protect\citeauthoryear{Jonas}{Jonas}{2016}]{Justin}
Jonas J.~L.,  2016, \mn@doi [.] {.}, .

\bibitem[\protect\citeauthoryear{{Junklewitz}, {Bell}, {Selig}  \&
  {En{\ss}lin}}{{Junklewitz} et~al.}{2016}]{resolve}
{Junklewitz} H.,  {Bell} M.~R.,  {Selig} M.,   {En{\ss}lin} T.~A.,  2016,
  \mn@doi [\aap] {10.1051/0004-6361/201323094}, \href
  {https://ui.adsabs.harvard.edu/abs/2016A&A...586A..76J} {586, A76}

\bibitem[\protect\citeauthoryear{Kass \& Raftery}{Kass \&
  Raftery}{1995}]{kassrafferty}
Kass R.~E.,  Raftery A.~E.,  1995, Journal of the American Statistical
  Association, 90, 773

\bibitem[\protect\citeauthoryear{{Law} et~al.,}{{Law} et~al.}{2011}]{law2011}
{Law} C.~J.,  et~al., 2011, \mn@doi [\apj] {10.1088/0004-637X/728/1/57}, \href
  {https://ui.adsabs.harvard.edu/abs/2011ApJ...728...57L} {728, 57}

\bibitem[\protect\citeauthoryear{{Leike} \& {En{\ss}lin}}{{Leike} \&
  {En{\ss}lin}}{2019}]{ensslinfromert}
{Leike} R.~H.,  {En{\ss}lin} T.~A.,  2019, \mn@doi [\aap]
  {10.1051/0004-6361/201935093}, \href
  {https://ui.adsabs.harvard.edu/abs/2019A&A...631A..32L} {631, A32}

\bibitem[\protect\citeauthoryear{{Littlefair}, {Burningham}  \&
  {Helling}}{{Littlefair} et~al.}{2017}]{2017MNRAS.466.4250L}
{Littlefair} S.~P.,  {Burningham} B.,   {Helling} C.,  2017, \mn@doi [\mnras]
  {10.1093/mnras/stw3376}, \href
  {https://ui.adsabs.harvard.edu/abs/2017MNRAS.466.4250L} {466, 4250}

\bibitem[\protect\citeauthoryear{{Macquart}, {Ekers}, {Feain}  \&
  {Johnston-Hollitt}}{{Macquart} et~al.}{2012}]{JP2012}
{Macquart} J.~P.,  {Ekers} R.~D.,  {Feain} I.,   {Johnston-Hollitt} M.,  2012,
  \mn@doi [\apj] {10.1088/0004-637X/750/2/139}, \href
  {https://ui.adsabs.harvard.edu/abs/2012ApJ...750..139M} {750, 139}

\bibitem[\protect\citeauthoryear{{Mauch} et~al.,}{{Mauch} et~al.}{2020}]{deep2}
{Mauch} T.,  et~al., 2020, \mn@doi [\apj] {10.3847/1538-4357/ab5d2d}, \href
  {https://ui.adsabs.harvard.edu/abs/2020ApJ...888...61M} {888, 61}

\bibitem[\protect\citeauthoryear{McAllister et~al.,}{McAllister
  et~al.}{2016}]{10.1093/mnras/stw2417}
McAllister M.~J.,  et~al., 2016, \mn@doi [Monthly Notices of the Royal
  Astronomical Society] {10.1093/mnras/stw2417}, 464, 1353

\bibitem[\protect\citeauthoryear{{Mertens}, {Ghosh}  \& {Koopmans}}{{Mertens}
  et~al.}{2018}]{mertens}
{Mertens} F.~G.,  {Ghosh} A.,   {Koopmans} L.~V.~E.,  2018, \mn@doi [\mnras]
  {10.1093/mnras/sty1207}, \href
  {https://ui.adsabs.harvard.edu/abs/2018MNRAS.478.3640M} {478, 3640}

\bibitem[\protect\citeauthoryear{O'Sullivan et~al.,}{O'Sullivan
  et~al.}{2012}]{10.1111/j.1365-2966.2012.20554.x}
O'Sullivan S.~P.,  et~al., 2012, \mn@doi [Monthly Notices of the Royal
  Astronomical Society] {10.1111/j.1365-2966.2012.20554.x}, 421, 3300

\bibitem[\protect\citeauthoryear{Pratley \& Johnston-Hollitt}{Pratley \&
  Johnston-Hollitt}{2020}]{pratley2020a}
Pratley L.,  Johnston-Hollitt M.,  2020, \mn@doi [The Astrophysical Journal]
  {10.3847/1538-4357/ab6e64}, 894, 38

\bibitem[\protect\citeauthoryear{{Pratley}, {Johnston-Hollitt}  \&
  {Gaensler}}{{Pratley} et~al.}{2020}]{pratley2020}
{Pratley} L.,  {Johnston-Hollitt} M.,   {Gaensler} B.~M.,  2020, arXiv
  e-prints, \href {https://ui.adsabs.harvard.edu/abs/2020arXiv201007932P} {p.
  arXiv:2010.07932}

\bibitem[\protect\citeauthoryear{Rajpaul, Aigrain, Osborne, Reece  \&
  Roberts}{Rajpaul et~al.}{2015}]{10.1093/mnras/stv1428}
Rajpaul V.,  Aigrain S.,  Osborne M.~A.,  Reece S.,   Roberts S.,  2015,
  \mn@doi [Monthly Notices of the Royal Astronomical Society]
  {10.1093/mnras/stv1428}, 452, 2269

\bibitem[\protect\citeauthoryear{{Rajpaul}, {Aigrain}  \& {Roberts}}{{Rajpaul}
  et~al.}{2016}]{2016MNRAS.456L...6R}
{Rajpaul} V.,  {Aigrain} S.,   {Roberts} S.,  2016, \mn@doi [\mnras]
  {10.1093/mnrasl/slv164}, \href
  {https://ui.adsabs.harvard.edu/abs/2016MNRAS.456L...6R} {456, L6}

\bibitem[\protect\citeauthoryear{Rasmussen \& Williams}{Rasmussen \&
  Williams}{2006}]{3569}
Rasmussen C.,  Williams C.,  2006, Gaussian Processes for Machine Learning.
Adaptive Computation and Machine Learning, MIT Press, Cambridge, MA, USA

\bibitem[\protect\citeauthoryear{Roberts, Osborne, Ebden, Reece, Gibson  \&
  Aigrain}{Roberts et~al.}{2013}]{article}
Roberts S.,  Osborne M.,  Ebden M.,  Reece S.,  Gibson N.,   Aigrain S.,  2013,
  \mn@doi [Philosophical transactions. Series A, Mathematical, physical, and
  engineering sciences] {10.1098/rsta.2011.0550}, 371, 20110550

\bibitem[\protect\citeauthoryear{{Schwarz}}{{Schwarz}}{1978}]{1978AnSta...6..461S}
{Schwarz} G.,  1978, Annals of Statistics, \href
  {https://ui.adsabs.harvard.edu/abs/1978AnSta...6..461S} {6, 461}

\bibitem[\protect\citeauthoryear{{Seaman}, {Galati}, {Jackson}  \&
  {Carlin}}{{Seaman} et~al.}{2013}]{MAR}
{Seaman} S.,  {Galati} J.,  {Jackson} D.,   {Carlin} J.,  2013, arXiv e-prints,
  \href {https://ui.adsabs.harvard.edu/abs/2013arXiv1306.2812S} {p.
  arXiv:1306.2812}

\bibitem[\protect\citeauthoryear{{Stil}, {Keller}, {George}  \&
  {Taylor}}{{Stil} et~al.}{2014}]{Stil_2014}
{Stil} J.~M.,  {Keller} B.~W.,  {George} S.~J.,   {Taylor} A.~R.,  2014,
  \mn@doi [\apj] {10.1088/0004-637X/787/2/99}, \href
  {https://ui.adsabs.harvard.edu/abs/2014ApJ...787...99S} {787, 99}

\bibitem[\protect\citeauthoryear{Sun et~al.,}{Sun et~al.}{2015}]{Sun_2015}
Sun X.~H.,  et~al., 2015, \mn@doi [The Astronomical Journal]
  {10.1088/0004-6256/149/2/60}, 149, 60

\bibitem[\protect\citeauthoryear{{Sun}, {Zhang}, {Wang}, {Zeng}, {Li}  \&
  {Grosse}}{{Sun} et~al.}{2018}]{2018arXiv180604326S}
{Sun} S.,  {Zhang} G.,  {Wang} C.,  {Zeng} W.,  {Li} J.,   {Grosse} R.,  2018,
  arXiv e-prints, \href {https://ui.adsabs.harvard.edu/abs/2018arXiv180604326S}
  {p. arXiv:1806.04326}

\bibitem[\protect\citeauthoryear{Taylor \& Jarvis}{Taylor \&
  Jarvis}{2017}]{Taylor_2017}
Taylor A.~R.,  Jarvis M.,  2017, \mn@doi [{IOP} Conference Series: Materials
  Science and Engineering] {10.1088/1757-899x/198/1/012014}, 198, 012014

\bibitem[\protect\citeauthoryear{{Van Eck} et~al.,}{{Van Eck}
  et~al.}{2018}]{lofarpyrmsynth}
{Van Eck} C.~L.,  et~al., 2018, \mn@doi [\aap] {10.1051/0004-6361/201732228},
  \href {https://ui.adsabs.harvard.edu/abs/2018A&A...613A..58V} {613, A58}

\bibitem[\protect\citeauthoryear{Wandelt \& Hansen}{Wandelt \&
  Hansen}{2003}]{PhysRevD.67.023001}
Wandelt B.~D.,  Hansen F.~K.,  2003, \mn@doi [Phys. Rev. D]
  {10.1103/PhysRevD.67.023001}, 67, 023001

\bibitem[\protect\citeauthoryear{Way, Foster, Gazis  \& Srivastava}{Way
  et~al.}{2009}]{wa06300n}
Way M.~J.,  Foster L.~V.,  Gazis P.~R.,   Srivastava A.~N.,  2009, \mn@doi
  [Astrophys. J.] {10.1088/0004-637X/706/1/623}, 706, 623

\makeatother
\end{thebibliography}




\bsp	
\label{lastpage}
\end{document}